\documentclass[headsepline=true, abstracton]{scrartcl}

\usepackage[utf8]{inputenc}

\usepackage{amssymb}
\usepackage{amsmath}
\usepackage{amsthm}
\usepackage{bm}

\usepackage{natbib}
\usepackage{hyperref}
\hypersetup{colorlinks=true, citecolor=darkgray, linkcolor=darkgray, urlcolor=darkgray}

\usepackage[moderate]{savetrees}

\usepackage[table,xcdraw]{xcolor}

\usepackage{graphicx}

 \usepackage{float}

\usepackage{setspace}

\usepackage{url}

\usepackage{geometry}

\begin{document}

\renewcommand{\refname}{Bibliography}

\onehalfspacing
\setlength{\headsep}{15mm}

\thispagestyle{plain}

\title{\Large Generative Dynamics of Supreme Court Citations: \\ Analysis with a New Statistical Network Model\thanks{
		This work was supported in part by NSF grants SES-1558661, SES-1637089, SES-1619644, and CISE-1320219. We thank Rachael Hinkle, Benjamin Kassow, Lauren Santoro, and Michael Nelson for helpful feedback on this project. The replication code and data for this paper can be found at \citet{replicationdata}.}}

\author{%
  Christian S. Schmid\footnote{Department of Statistics, The Pennsylvania State University, schmid@psu.edu}%
  \and Ted Hsuan Yun Chen\footnote{Faculty of Social Sciences, University of Helsinki, ted.hsuanyun.chen@gmail.com}%
   \and Bruce A. Desmarais\footnote{Department of Political Science, The Pennsylvania State University, bdesmarais@psu.edu}%
  }

\date{Conditionally Accepted at \textit{Political Analysis}\\ December 29, 2020}

\maketitle
\begin{abstract}
\noindent The significance and influence of US Supreme Court majority opinions derive in large part from opinions' roles as precedents for future opinions. A growing body of literature seeks to understand what drives the use of opinions as precedents through the study of Supreme Court case citation patterns. We raise two limitations of existing work on Supreme Court citations. First, dyadic citations are typically aggregated to the case level before they are analyzed. Second, citations are treated as if they arise independently. We present a methodology for studying citations between Supreme Court opinions at the dyadic level, as a network, that overcomes these limitations. This methodology---the citation exponential random graph model, for which we provide user-friendly software---enables researchers to account for the effects of case characteristics and complex forms of network dependence in citation formation. We then analyze a network that includes all Supreme Court cases decided between 1950 and 2015. We find evidence for dependence processes, including reciprocity, transitivity, and popularity. The dependence effects are as substantively and statistically significant as the effects of exogenous covariates, indicating that models of Supreme Court citation should incorporate both the effects of case characteristics and the structure of past citations.
\end{abstract}

\onehalfspacing
\section{Introduction}
\label{sec:overview}

United States Supreme Court opinions exercise authority and influence, in part, through their roles as precedents affecting future jurisprudence in the US. The findings regarding the nature of the influences of precedent on the Supreme Court have been mixed, but the balance of the literature finds that past decisions exert some form of influence on the justices' decision making \citep{knight1996norm,gillman2001s,richards2002jurisprudential,hansford2006politics,bailey2008does,bailey2011constrained,pang2012endogenous}. Despite a considerable body of research that focuses on how precedents shape decision making on the Court, relatively little work has focused on understanding which past opinions are cited by an opinion. Our focus in this paper is to provide what is, to our knowledge, the first comprehensive analysis of exactly which cases are cited in an opinion. We follow an emerging body of work on legal citations, and treat the system of citations as a network \citep[e.g., ][]{caldeira1988legal,fowler2007network, fowler2008authority,bommarito2009law,lupu2012precedent,pelc2014politics,ethayarajh2018rose}. 

We are not the first to ask what predicts the citations in US Supreme Court Opinions. Indeed, a voluminous body of work has sought to explain how many times an opinion is cited \citep[e.g.,][]{cross2010determinants,benjamin2012standing,fix2019effect}, when an opinion is cited \citep[e.g.,][]{black2013citation,spriggs2001explaining}, and how many cases are cited by an opinion \citep[e.g.,][]{lupu2013strategic}---all focused on the US Supreme Court. One common feature of the research design in all of these studies is that the observations are at the case or case-year level. The outcome variables in these analyses are defined as measures of the number of citations to a case over a period of time, the number of citations to a case at a particular time, or a measurement on the cases cited by a case. These are case-level studies in that, based on the unit of analysis, it is impossible to determine both the origin and target case of a citation that contributes to the dependent variable.

An alternative approach to case-level analysis of citations would be to model them in the directed dyadic form through which they arise. A case decided at time $t$ can cite (or not cite) each case decided previously, and in the US Supreme Court, each other case decided at time $t$. We are aware of one prior study, \citet{clark2010locating}, in which a statistical model is used to analyze directed dyadic citations between cases. However, \citet{clark2010locating} use a dyadic latent variable model in order to estimate ideal points for Supreme Court Opinions, but do not use this model to understand the relationships between explanatory variables and the formation of citation ties between opinions. We build upon the literature on citation analysis both methodologically and substantively. Methodologically, we develop a novel extension of a statistical model for networks, which we adapt to the network structural constraints of court citations. Second, we apply this methodology to a half-century of directed dyadic citations between U.S. Supreme Court opinions.

There are two benefits of analyzing citations at the directed dyad level. The first is that directed dyadic analyses can test both dyadic and case-level hypotheses. For example, case-level analyses can model whether opinions supported by a liberal majority coalition are more likely than those supported by a conservative majority coalition to be cited heavily in the future, but they cannot precisely model whether liberal cases will be cited more by liberal cases than by conservative cases. Thus, the first reason for analyzing citations at the dyadic level is to expand the set of hypotheses that can be tested. The second reason for studying citations at the directed dyadic level is that, as articulated in the growing literature on legal citation networks, citations form complex networks in which a citation at one point in time may influence future citations. This phenomenon of complex dependence is very common in networks of many types, but processes specific to Supreme Court citations create interdependence in citations. For example, if opinion $i$ relies heavily on opinion $j$ as precedent, opinion $i$ is likely to discuss the legal basis for opinion $j$, and as a consequence, cite some of the opinions cited by opinion $j$. Suppose opinion $k$ is cited by opinion $j$. Opinion $k$ is more likely to be cited by opinion $i$ because opinion $i$ relies heavily on $j$, and opinion $j$ cites $k$.  This is a special case of a very common process on networks referred to as ``triad closure''. Complex dependence is theoretically interesting on its own merits, but the effects of covariates cannot be reliably identified---either in terms of coefficient values or standard errors---without accounting for the interdependence inherent in networks \citep{cranmer2016critique}. 

In this paper we develop a theoretical case that citations on the US Supreme Court are characterized by forms of complex dependence that are common in networks. We then develop an extension of a model---the exponential random graph model (ERGM)---that can incorporate both exogenous covariates and complex forms of interdependence into a directed dyadic analysis of citations. Finally, we develop and estimate a specification of this model in an analysis of US Supreme Court citations between 1950 and 2015. We find robust support for the inherent complexity underpinning the formation of citation ties, and show that incorporating complex dependence into the model of citation formation significantly improves the model's fit.

We offer three contributions. First, we advance our understanding of the factors that drive citations between U.S. Supreme Court opinions. Second, for those who study judicial citations in general (e.g., U.S. state supreme courts \citep{hinkle2016transmission}, international courts \citep{lupu2012precedent},  German lower courts \citep{berlemann2020disposition}), we illustrate network-theoretic considerations that are likely to apply beyond the context of Supreme Court opinions. Third, we offer a novel extension of a statistical model for networks, and disseminate this model as a package for the R statistical software \citep{cergm}, which can be used for any form of citations. For example, patent citations are used to measure both causes and consequences of innovation in the field of political economy \citep{akcigit2018growth,dincer2019does}, citations in academic journal articles and syllabi have been used to study gender bias in political science  \citep{dion2018gendered,maliniak2013gender,hardt2019gender,atchison2017negating}, and citations in documents produced in the policymaking process have been used to study links between public policy and scientific expertise \citep{costa2016science,koontz2018use,pattyn2020knowledge}. Our contributions are both substantive and methodological. We focus most heavily on the first---the study of the U.S. Supreme Court citation network, since we see it as the most substantial innovation with respect to the existing literature that we offer in the current paper.

\section{Network Processes in Supreme Court Citations} 

When it comes to the development and testing of theory, the defining feature of networks is that the micro-level unit of analysis---the relationship between two entities (i.e., the citation from one opinion to another) is a component of a complex system of relations. The formation (or lack thereof) of that relationship cannot be fully understood without considering how the relationship fits into the system. Analytical designs that account only for covariates in explaining tie formation are incomplete theoretically, and, as a consequence, are subject to a form of omitted variable bias \citep{cranmer2016critique}. Citations in legal opinions are unique in terms of the windows into network dependencies offered by the texts of the opinions. A number of common structural dependencies that are found in networks are likely to apply to citations in Supreme Court opinions. In this section we present these dependence forms, and document the mechanisms by which they arise through archetypal passages in example opinions. 

We should note that we do not distinguish between positive and negative citations in this theoretical framework. The dynamics we outline are not specific to a particular type of citation. The only distinction we draw (in the empirical analysis) is the instance in which a case has been overruled. In the extreme instance of overruled precedents, we assume that (and test whether) a case is much less likely to be cited after it has been overruled.

The first network property that we theorize in the context of Supreme Court citations is transitivity. In a network of directed relations (e.g., $i$ cites $j$, but $j$ doesn't cite $i$) transitivity refers to the tendency for $i$ to send a tie to $k$ if $i$ sends a tie to $j$ and $j$ sends a tie to $k$ \citep{holland1971transitivity,hallinan1990sex}. In undirected networks, transitivity is simply the process by which friends of friends become friends (i.e., a friend of a friend is a friend). The term, ``transitive closure'' refers to a tie forming from $i$ to $k$ in response to extant ties from $i$ to $j$ and $j$ to $k$. When writing opinions, Supreme Court justices present the legal bases for their rulings, which often involves discussing the most primary/relevant precedents underpinning these legal bases, but also the precedents and legal rules on which the primary precedents were based. This process of presenting several layers/levels of precedent in an opinion follows the structure of transitive closure exactly---opinion $i$ cites opinion $j$ as a primary precedent, and then cites opinion $k$ because opinion $j$ cites opinion $k$. The two examples presented below illustrate this process.

\begin{figure}[bt]
	\centering
	\includegraphics[width = 0.49\textwidth,trim= 2cm 2cm 3cm 2cm,clip=true ]{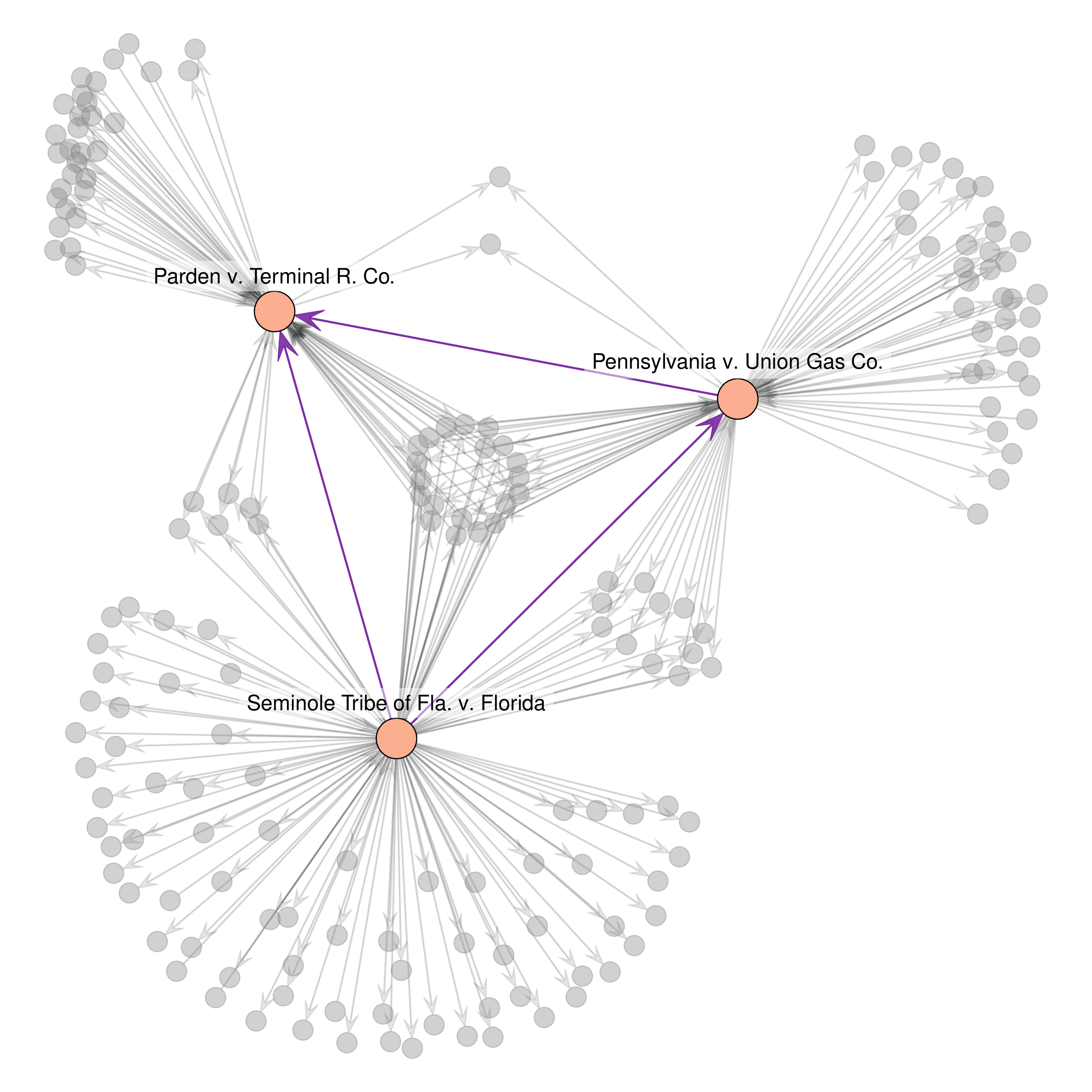}\hspace{0.2cm}%
	\includegraphics[width = 0.49\textwidth,trim= 2cm 2cm 3cm 2cm,clip=true ]{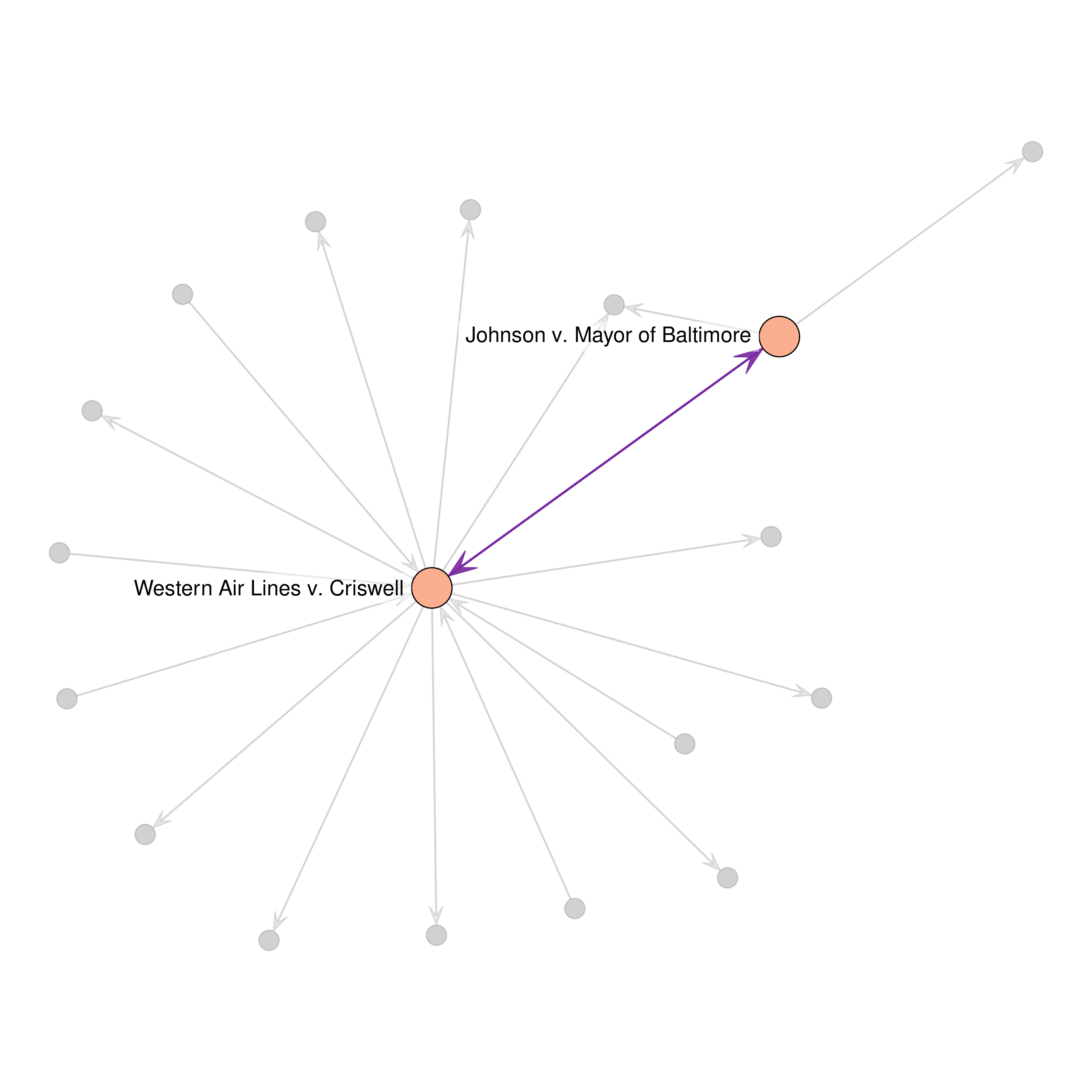}
	\caption{Illustrations of transitive triangle connecting US Supreme Court opinions through citations (left) and a reciprocal tie between two US Supreme Court opinions (right).}
	\label{fig:transitivity}
\end{figure}

In the first example, a passage from Kansas v. Marsh (548 U.S. 163, 2006)---a case considering the constitutionality of a death sentence statute in Kansas. In this example, the case Stringer v. Black is cited by Kansas v. Marsh as a case that is quoted by Sochor v. Florida. The primary precedent under discussion in this passage of the opinion is Sochor v. Florida, but Stringer v. Black is cited as a result of its role in the Sochor v. Florida opinion.
\begin{quotation}
	The statute thus addresses the risk of a morally unjustifiable death sentence, not by minimizing it as precedent unmistakably requires, but by guaranteeing that in equipoise cases the risk will be realized, by ``placing a `thumb [on] death's side of the scale,' '' Sochor v. Florida, 504 U. S. 527, 532 (1992) (quoting Stringer v. Black, 503 U. S. 222, 232 (1992); alteration in original).
\end{quotation} 
The second example, which we illustrate visually in Figure \ref{fig:transitivity} is a passage from Seminole Tribe of Fla. v. Florida, 
(517 U.S. 44 1996)---a case addressing the rights of groups and citizens to sue states in federal court. In this example, Pennsylvania v. Union Gas Co
(491 U.S. 1, 1989) is the primary precedent being critiqued, and several cases are cited and discussed in terms of their roles as precedents in the Union Gas opinion. We highlight one---Parden v. Terminal R. Co., which is cited and discussed in both the Seminole and Union Gas opinions. 
\begin{quotation}
	Never before the decision in Union Gas had we suggested that the bounds of Article III could be expanded by Congress operating pursuant to any constitutional provision other than the Fourteenth Amendment. Indeed, it had seemed fundamental that Congress could not expand the jurisdiction of the federal courts beyond the bounds of Article III. Marbury v. Madison, 1 Cranch 137 (1803). The plurality's citation of prior decisions for support was based upon what we believe to be a misreading of precedent. See Union Gas, 491 U. S., at 40-41 (SCALIA, J., dissenting). The plurality claimed support for its decision from a case holding the unremarkable, and completely unrelated, proposition that the States may waive their sovereign immunity, see id., at 14-15 (citing Parden v. Terminal Railway of Ala. Docks Dept., 377 U. S. 184 (1964)), and cited as precedent propositions that had been merely assumed for the sake of argument in earlier cases, see 491 U. S., at 15 (citing Welch v. Texas Dept. of Highways and Public Transp., 483 U. S., at 475-476, and n. 5, and County of Oneida v. Oneida Indian Nation of N. Y., 470 U. S., at 252).'
\end{quotation}

The second network property that we hypothesize is reciprocity \citep{erikson2013formalist} among cases that are decided in the same term. We expect that opinions that are written within the same term, and cover highly similar cases will cite each other. The first case in our example reciprocal dyad is a passage from Western Air Lines v. Criswell (472 U.S. 400, 1985)---a case considering mandatory retirement in the context of age discrimination laws. The second case in the dyad, Johnson v. Mayor of Baltimore (472 U.S. 353, 1985) is another case considering whether mandatory retirement violates the Age Discrimination in Employment Act. The mutual edge connecting these two cases is visualized in Figure \ref{fig:transitivity}. These cases addressed very similar legal questions, which increased the likelihood that they would inform each other, and the opinions were written within the same term, which made it possible for them to cite each other. 
\begin{quotation}
	{\em From Western Air Lines:} On a more specific level, Western argues that flight engineers must meet the same stringent qualifications as pilots, and that it was therefore quite logical to extend to flight engineers the FAA's age 60 retirement rule for pilots. Although the FAA's rule for pilots, adopted for safety reasons, is relevant evidence in the airline's BFOQ defense, it is not to be accorded conclusive weight. Johnson v. Mayor and City Council of Baltimore, ante at 472 U. S. 370-371. The extent to which the rule is probative varies with the weight of the evidence supporting its safety rationale and "the congruity between the . . . occupations at issue." Ante at 472 U. S. 371. In this case, the evidence clearly established that the FAA, Western, and other airlines all recognized that the qualifications for a flight engineer were less rigorous than those required for a pilot.
	\\~\\
	{\em From Johnson:} The city, supported by several amici, argues for affirmance nonetheless. It asserts first that the federal civil service statute is not just a federal retirement provision unrelated to the ADEA, but in fact establishes age as a BFOQ for federal firefighters based on factors that properly go into that determination under the ADEA, see Western Air Lines, Inc. v. Criswell, post p. 472 U. S. 400. Second, the city asserts, a congressional finding that age is a BFOQ for a certain occupation is dispositive of that determination with respect to nonfederal employees in that occupation. 
\end{quotation} 

\begin{figure}[bt]
	\centering
	\includegraphics[width = 0.8\textwidth,trim= 1cm 4cm 1cm 3cm,clip=true ]{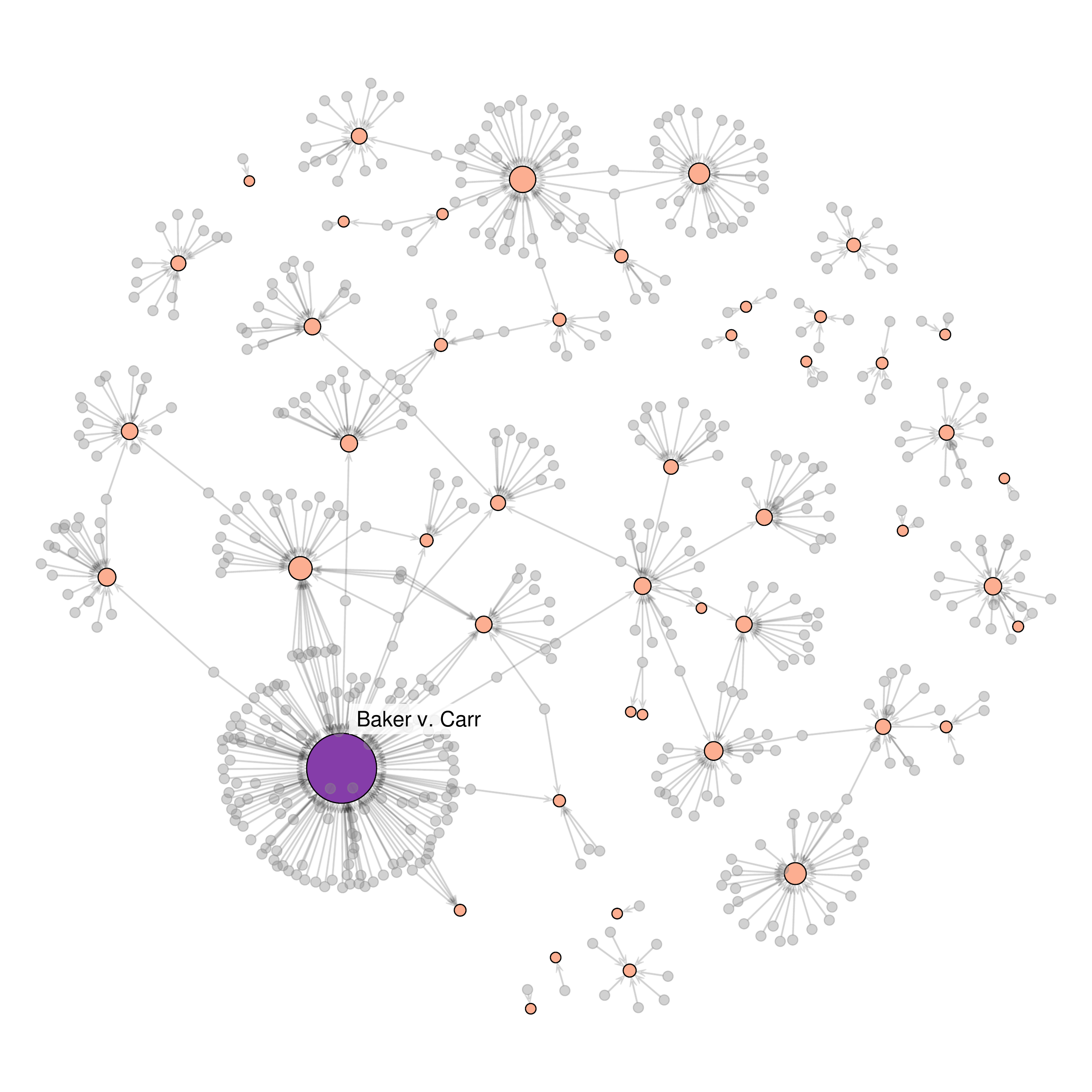}
	\caption{Illustration of ties sent to a landmark Supreme Court opinion via citations.}
	\label{fig:popularity}
\end{figure}

The third, and final, network property we consider in the context of Supreme Court citations is popularity. Popularity, also termed ``preferential attachment'' is the tendency for ties to be sent to nodes to which many ties have already been sent \citep{barabasi1999emergence,chayes2013mathematics}. Citations to an opinion signal both the Court's awareness of the legal reasoning of the case and the Court's evaluation that the opinion is an authoritative precedent. The more citations, the stronger this signal. Landmark cases, or those that establish new legal rules, are particularly authoritative and accrue citations from most future opinions that follow the respective line of reasoning. The passage below, from Oregon v. Mitchell (400 U.S. 112, 1970)---a case on the legality of state age restrictions on voting in federal elections---illustrates this popularity dynamic. In this opinion passage Baker v. Carr is cited in reference to its role as a landmark precedent, and noted for the number of other cases by which it has been followed. and for which an authoritative opinion is referenced, and even discussed in terms of the number of other cases by which it was followed. The language in this passage suggests that the attention to Baker v. Carr in previous Court opinions is in part responsible for its authority in Oregon v. Mitchell. The citations to Baker v. Carr are visualized in Figure \ref{fig:popularity}.
\begin{quotation}
	The first case in which this Court struck down a statute under the Equal Protection Clause of the Fourteenth Amendment was Strauder v. West Virginia, 100 U. S. 303, decided in the 1879 Term. [Footnote 2/1] In the 1961 Term, we squarely held that the manner of apportionment of members of a state legislature raised a justiciable question under the Equal Protection Clause, Baker v. Carr, 369 U. S. 186. That case was followed by numerous others, e.g.: that one person could not be given twice or 10 time the voting power of another person in a state-wide election merely because he lived in a rural area..."
\end{quotation} 

These three properties---transitivity, reciprocity, and popularity---form the core of our network theory of U.S. Supreme Court citations. We also seek to model the effects of covariates (i.e., case features) on citation formation.  Next we introduce a modeling framework that can incorporate all of these effects into a single model, and adapt to the structural constraints of citation networks.

\section{The Citation Exponential Random Graph Model}
We develop a methodology that can be used to jointly test for the effects of covariates on citations---as have been studied in prior research, and test for dependence effects, as we hypothesize above. To accomplish this, we extend a model that has been developed to jointly represent covariate and dependence effects in network data, and has seen extensive application in recent political networks research---the exponential random graph model (ERGM) \citep[e.g.,][]{bratton2011networks, box2014evolution,duque2018recognizing,osei2018elite}. The network structures for which ERGMs are currently designed are insufficient to account for the structure of citation networks.

\begin{figure}[bt]
	\centering
	\includegraphics[width = 0.775\textwidth ]{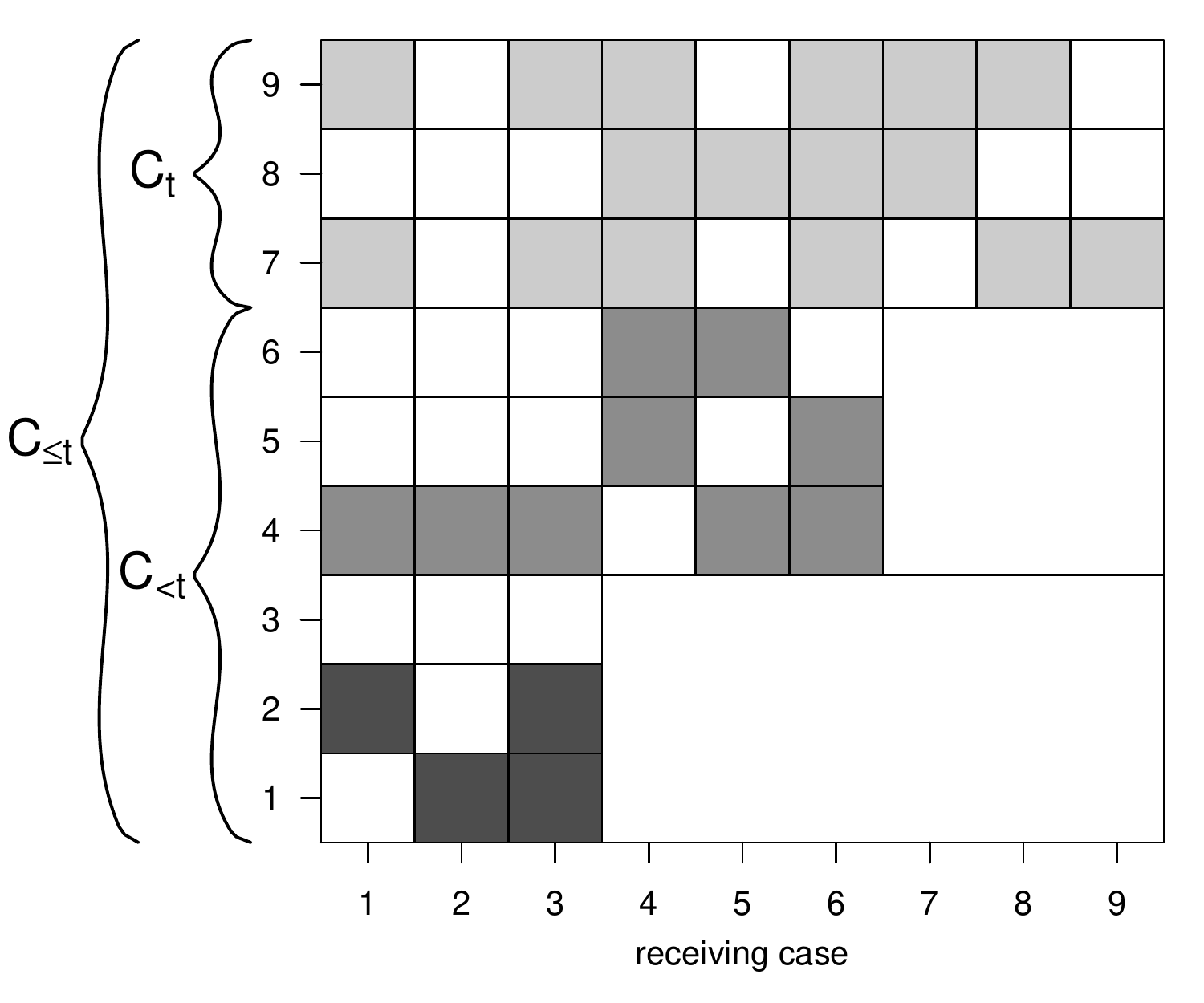}
	\caption{Illustration of temporal structure of the Supreme Court Citation Network. $C_{\leq t}$ is the entire set of citations (and non-citations) on which citations and non-citations at time $t$ (i.e., $C_t$) depend. $C_t$ are conditioned on the citations and non-citations established before time $t$ (i.e., $C_{< t}$). The shaded small squares are hypothetical observed citations, and the white small squares are citations that could have been observed but were not. The regions of the matrix that are represented by large white rectangles are citations that could not have been observed since the citing case would have been decided in a term that preceded the term of the cited case. The citing case ID is given in the row and the prospective cited case is given in the column. }
	\label{fig:ctergm}
\end{figure}

We develop the citation ERGM (cERGM), to account for the structural constraints that apply to the network of Supreme Court citations. These structural constraints amount to three departures from the structure of networks for which ERGMs are currently designed. First, the citation network is partially acyclic. If two cases are decided during the same term, they can cite each other, forming a mutual edge (or two-cycle). However, if case $i$ is decided before case $j$, case $j$ can cite case $i$, but case $i$ cannot cite case $j$. Second, new edges can be created over time, but cannot be eliminated. Unlike in, e.g., an alliance network, in which two countries can dissolve an alliance, once a citation exists in a citation network it cannot be dissolved. Third, the set of nodes in the network must increase for new edges to be created. In conventional ERGMs, the number of nodes in the network can increase or decrease in each time period, and is typically stable over time. In a citation network, new edges (and non-edges) are introduced over time via the introduction of new nodes. The structure of the citation network is depicted in Figure \ref{fig:ctergm}---a hypothetical citation network established over three time periods, with three cases decided in each time period. We denote $C_t$ to be the set of citation and non-citations added to the network at time $t$ (i.e., via the addition of three cases), $C_{ <t}$ to be the citations among cases decided before time $t$, and $C_{ \leq t}$ to denote the entire set of citations and non-citations on which $C_t$ can depend through the cERGM specification.

The likelihood function of the cERGM is given by
\begin{equation}
l(\bm{\theta},C_{\leq t}) =  \frac{ \exp \left[ {\bm{\theta}'\bm{h}(C_{t},C_{<t}) } \right] }{ \sum_{C_t^* \in \mathcal{C}_t} \exp \left[ {\bm{\theta}'\bm{h}(C^*_{t},C_{<t}) }\right]  },
\end{equation}
where $\bm{\theta}$ is a vector of real-valued parameters, and $\bm{h}(C_{t},C_{<t})$ is a vector of scalar-valued functions that each quantify a feature of the citation network (e.g., the relationship between citation ties and a case attribute, the number of mutual edges in the citation network). $\exp \left[ {\bm{\theta}'\bm{h}(C_{t},C_{<t}) } \right]$ is a positive weight that is proportional to the probability of observing any particular form of the citations and non-citations added to the network at time $t$.  The denominator of Equation 1 represents a normalizing constant, in which the positive weight is summed over all possible configurations of $C_{t}$, from the network in which the cases added to the network at time $t$ send no citations at all, to the network in which the cases added at time $t$ cite every possible case, and everything in between.

Though the likelihood function of the cERGM is quite different from those of conventional regression models, analysis of the conditional probability of a single citation from case $i$ to case $j$ reveals that we can interpret the parameters similar to logistic regression coefficients. $$ P(C_{ij,t} = 1 | C_{-ij,t}, C_{ < t}) = \frac{\exp \left[ {\bm{\theta}'\bm{h}(C_{t},C_{<t}| C_{ij,t} = 1) } \right]}{ \exp \left[ {\bm{\theta}'\bm{h}(C_{t},C_{<t}| C_{ij,t} = 1) } \right] + \exp \left[ {\bm{\theta}'\bm{h}(C_{t},C_{<t}| C_{ij,t} = 0) } \right]}, $$ $$ \text{~~~~~~~~~~~~~~~~~~~~~~~~~~~~~~~} = \frac{1}{ 1 + \exp \left[ - {\bm{\theta}'\left(\bm{h}(C_{t},C_{<t}| C_{ij,t} = 1) - \bm{h}(C_{t},C_{<t}| C_{ij,t} = 0)\right)} \right]}, $$
where $C_{ij,t} = 1$ indicates that case $i$ cites case $j$, $C_{ij,t} = 0$ indicates that case $i$ does not cite case $j$, $C_{-ij,t}$ is the observed elements of $C_{t}$ except $C_{ij,t}$, and $\left(\bm{h}(C_{t},C_{<t}| C_{ij,t} = 1) - \bm{h}(C_{t},C_{<t}| C_{ij,t} = 0)\right)$ is the change in $\bm{h}(C_{t},C_{<t})$ that results from toggling $C_{ij,t} = 0$ to $C_{ij,t} = 1$. This re-arrangement illustrates that the parameters can be interpreted in terms of the change in the log odds of a citation from $i$ to $j$ given a one-unit increase in the corresponding element of $\bm{h}$, conditional on the other citations observed in the network. For example, if the value of $\theta$ corresponding to an element of $\bm{h}$ that counts the number of mutual edges in the network is 0.5, then the log odds of observing $C_{ij,t} = 1$ increases by 0.5 if case $i$ is cited by case $j$ (as compared to the configuration in which case $j$ does not cite case $i$). The logit form conditional probability is well known for the ERGM family \citep{goodreau2009birds}.

\section{Empirical Analysis}
\begin{figure}[bt]
	\centering
	\begin{minipage}{.37\linewidth}
		\centering
		\includegraphics[width = 0.99\textwidth]{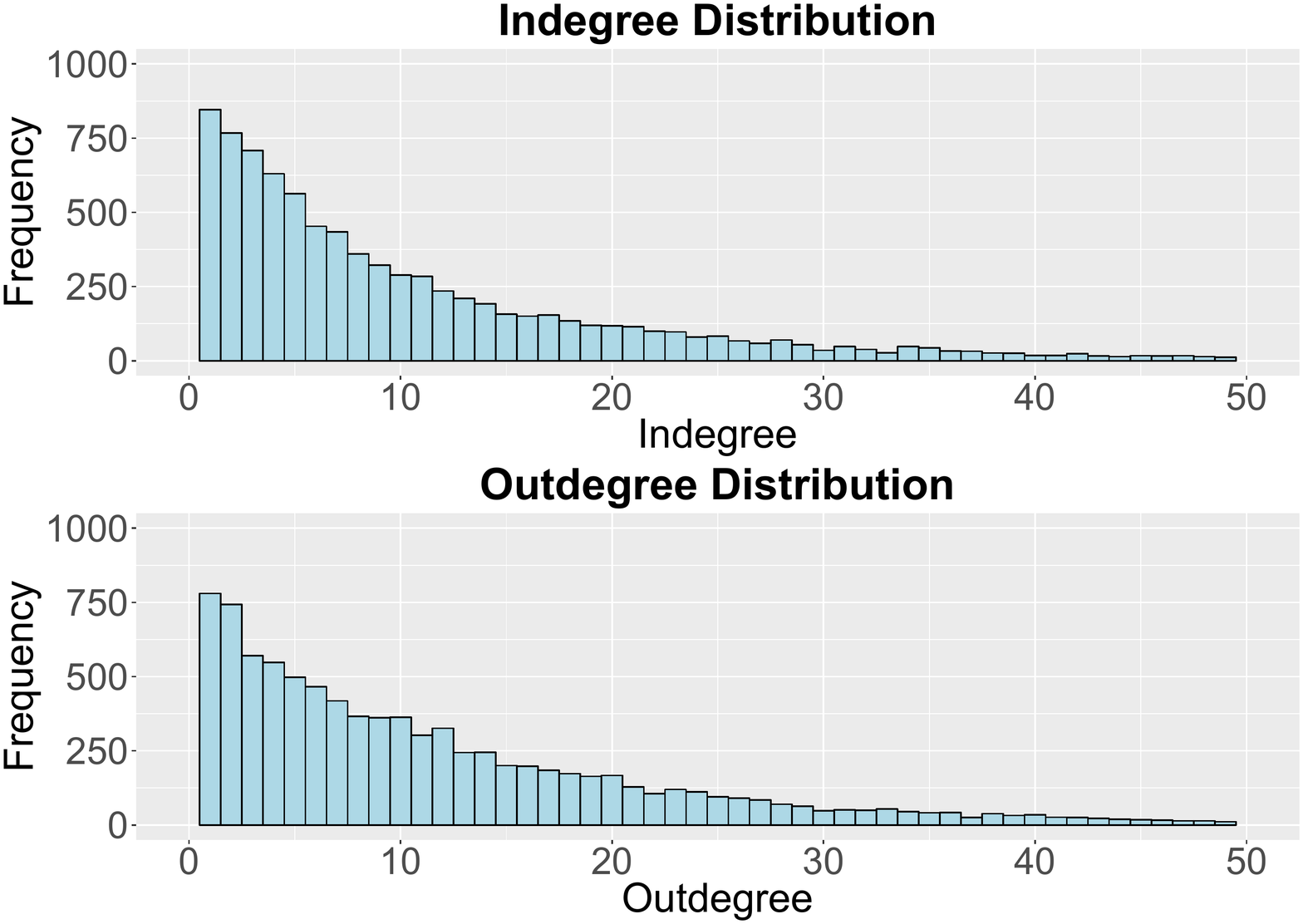}
		\includegraphics[width = 0.8\textwidth]{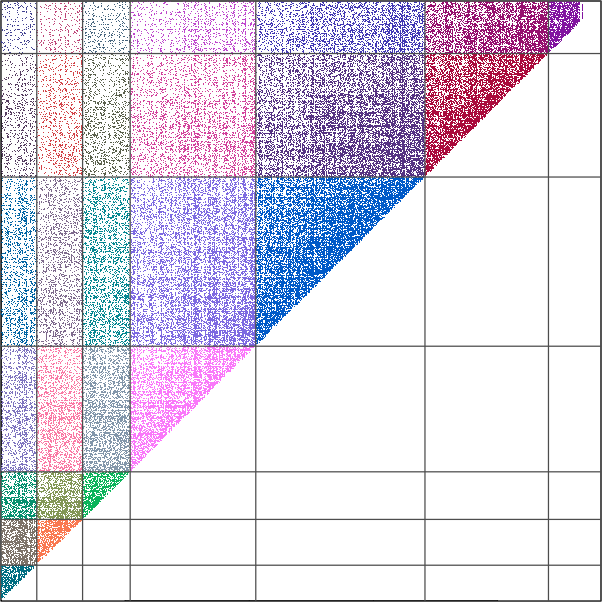}%
	\end{minipage}%
	\begin{minipage}{.63\linewidth}
		\includegraphics[width = \textwidth]{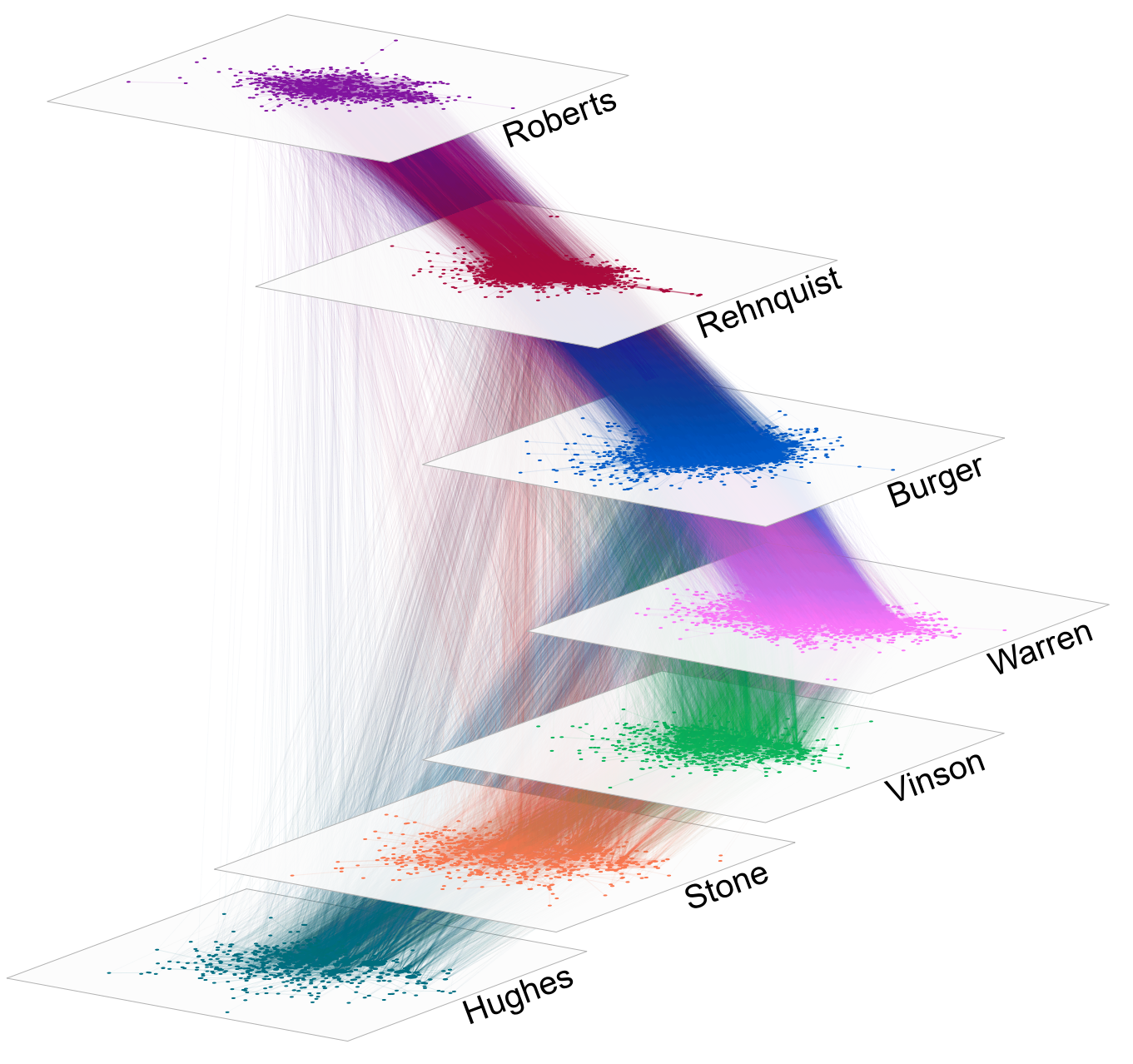}
	\end{minipage}
	\caption{Supreme Court Citation Network, 1937-2015. Network visualization on the right. Nodes are Supreme Court cases, color-coded based on the chief justice presiding over the court. On the top left is the in- and outdegree distribution of the network. There are cases with an in- or outdegree  $>$50, but they are not captured in this figure. The bottom left shows the citation data in adjacency matrix format following \autoref{fig:ctergm}.}
	\label{degree_dist}
\end{figure}

Our three data sources for this study include the Supreme Court Database (SCDB) \citep{spaeth2014supreme}, Martin-Quinn scores \citep{martin2002dynamic}, and Supreme Court citation data provided by the CourtListener Free Law Project \citep{CourtListener}. In the next section we explain the variables we construct using these data sources\footnote{Data and replication code are provided by \citet{replicationdata}}. We limit the Supreme Court terms included in our analysis to those that are covered by all three of these data source (1937--2015).\footnote{There were 145 cases that were listed in the SCDB but could not be matched to a case in the CourtListener data. We decided to exclude these 145 cases from our analysis. Additionally, since the most commonly used data on Supreme Court citations in political science come from \citet{fowler2007network}, we compared the Fowler data to the CourtListener data in the time interval that they overlap (1937--2001). We found considerable agreement---over 95\% of the citations recorded in the CourtListener dataset were also in the Fowler dataset, and over 96\% of the citations recorded in the Fowler data were also in the CourtListener dataset.} The Supreme Court citation network from $1937 - 2015$ consists of $10,020$ cases. The breakdown of the data by the Court's Chief Justice is presented in Table \ref{tab:chiefs}. The network has a total of $112,939$ citation ties.\footnote{In order to focus on citation actions that were not intended to totally invalidate an opinion, we exclude $315$ citations that caused the cited case to be overruled. The data on overruling citation came from \citet{senate2016constitution}. Our results are virtually unchanged if we include the overruling citations.} The in- and outdegree distributions (i.e., the distributions of the number of citations sent and received by cases, respectively), are visualized in Figure \ref{degree_dist}. The maximum indegree (i.e., number of cases citing to a case) is $230$ and the maximum outdegree (i.e., number of cases cited by a case) is $162$. The majority of cases cite to and/or are cited by twenty or fewer other cases.

\begin{table}[bt]
	\label{tab:chiefs}
	\centering
	\begin{tabular}{l l c r}
		 {Chief Justice} & {Terms} & {Total Number Cases} & {Cases/Term} \\ \hline
		CE Hughes*              & 1937 - 1941                                          & 628                                                               & 125.6                                                     \\ 
		HF Stone                & 1942 - 1946                                          & 756                                                               & 151.2                                                     \\ 
		FM Vinson               & 1946 - 1953                                          & 789                                                               & 98.63                                                      \\ 
		E Warren                & 1954 - 1969                                          & 2149                                                              & 126.41                                                     \\ 
		WE Burger               & 1970 - 1986                                          & 2805                                                             & 155.83                                                     \\ 
		W Rehnquist            & 1987 - 2005                                          & 2022                                                              & 106.42                                                      \\ 
		J Roberts **           & 2006 - 2015                                          & 871                                                              & 87.1                                           \\ 
	\end{tabular}
		\caption{For the time range of interest (1937 - 2015) this table displays the chief justices, the time range they served as chief justice, the number of cases in their time range as well as the average number of cases per year.\\\hspace{\textwidth} * CE Hughes served as chief justice from 1930 - 1941. \\\hspace{\textwidth} ** J Roberts still serves as chief justice (retrieved 5/2020).}
\end{table}

The degree distributions indicate that there is a long tail to both the number of citations sent and received. These long-tailed (i.e., high kurtosis) distributions provide preliminary evidence of substantial heterogeneity in the features that drive citations to and from cases \citep{strogatz2001exploring}. Figure \ref{degree_dist} displays the citation network network from 1937--2015. We see here that the densest rates of tie formation tend to be between consecutive courts (e.g., the Stone Court is much more tightly tied to the Hughes Court than the Rehnquist Court is to the Hughes Court). This pattern lends preliminary support to the hypothesis that the rate of citations to a case decreases over time.

We fit the cERGM separately for every Supreme Court term between 1950 and 2015, meaning that we have 66 models where the outcome variables are the citations sent during the given term. Our approach of fitting separate term-by-term models is motivated by the temporally-dynamic data generating process. First, as our results will show, the temporal context and justice composition of a court will lead to differences in how opinions are written and therefore how citations are used. Second, based on the findings of \citet{shalizi2013consistency}, we know that the set of parameter values that best fits a network should change as more nodes are added to the network. This means it would be inappropriate to assume a constant set of parameter values as more cases are added each term. Practically, our dataset is large enough to support this approach, with each term introducing the potential for hundreds-of-thousands of new citations. This gives us the statistical power necessary to estimate a separate set of parameter values for each term. Further, while the entire set of citations can in theory be fit in a single network with varying parameters, it is technically difficult. While the network for the 1950 term consists of a manageable $1,962$ nodes, the size of the network increases to $10,020$ nodes in the 2015 term, making estimation more challenging. 

Estimation of the cERGM parameters, which is done with a computationally-intensive, simulation-based approach, is presented in detail in the appendix. The basis of our computation is the \textbf{ergm}-package \citep{HunterDavidR..2008} in \textbf{R} \citep{RCore}. In addition, we developed a wrapper \textbf{R}-package called \textbf{cERGM} \citep{cergm} to conveniently fit the cERGM.

\subsection{cERGM specification}

The cERGM we estimate includes two classes of terms---one that captures the effects of covariates on tie formation, and another that captures the complex dependence processes that we expect to observe in the Supreme Court Citation Network. We describe our model specification by defining the terms within these two classes.

\subsubsection{Covariate terms}\label{covariate_terms}

Covariate effects are accounted for in the cERGM via the term that is used to specify the effect of covariates in other ERGM family models, as
$$h_{covariate}(C_t,C_{<t},X) =  \sum_{ij} C_{ij,t}X_{ij}.$$ Since $C_{ij,t}$ is a binary indicator of whether case $i$ cites case $j$, $h_{covariate}$ amounts to the sum of covariate values among directed dyads for which we observed a citation. The dyadic interpretation of the coefficient attributed to this term is the change in the log odds of a tie from $i$ to $j$ given a one-unit increase in $X_{ij}$ (i.e., exactly the interpretation of the effect of a covariate in logistic regression). We include several exogenous covariates based on this standard formulation. 

The first covariate we incorporate into the model accounts for the degree to which cases cite those that are similar in terms of the ideological positions of the justices who supported the decision. We account for this effect following \citet{spriggs2001explaining}, who find that cases are more likely to be overruled when the Court is ideologically distant from the median justice in the majority coalition that decided the case. \citet{clark2010locating} find that the majority opinion falls at the ideal point of the median member of the majority coalition in the case. We include a covariate term in which $X_{ij}$ is the absolute difference between the Martin-Quinn scores \citep{martin2002dynamic} of the median justices in the majority coalitions for cases $i$ and $j$. We expect this variable to have a negative effect, which would mean that cases cite those that are ideologically similar.

We include one covariate that accounts for the rate at which cases that fall under the same issue area cite each other. The variable $X_{ij}$ indicates whether cases $i$ and $j$ share the same issue area, e.g., $X_{ij}=1$ if cases $i$ and $j$ are classified into the same issue area, and $0$ otherwise. The Issue area data comes from the Supreme Court Database (SCDB) \citep{spaeth2014supreme}. We include these variables because \citet{cross2010determinants} finds that the number of citations to Supreme Court opinions depends heavily on the issue area of the case. Table \ref{issue_area_coding} reports the issue areas covered in our data.

We also include an indicator variable that models the rate at which justices cite themselves. Similar to the same issue area variable, this effect is modeled with a single indicator variable in which $X_{ij} = 1$ if the majority opinions in cases $i$ and $j$ were written by the same justice. The data is taken from the SCDB \citep{spaeth2014supreme}.

\begin{table}[bt]

	\label{issue_area_coding}
	\centering
	\begin{tabular}{llll}
		1 & Criminal Procedure & 8  & Economic Activity    \\
		2 & Civil Rights       & 9  & Judicial Power       \\
		3 & First Amendment    & 10 & Federalism           \\
		4 & Due Process        & 11 & Interstate Relations \\
		5 & Privacy            & 12 & Federal Taxation     \\
		6 & Attorneys          & 13 & Miscellaneous        \\
		7 & Unions             & 14 & Private Action      
	\end{tabular}
	\caption{Assigned numbers for the variable \textit{Issue Area}. This information originates from the Supreme Court Database code book.}
\end{table}

We model the way in which citations to a case depend upon the age of a case. For this we use a second-order polynomial in which one covariate $X_{ij}$ is defined as the age of case $j$ at the time that case $i$ is decided, and another term in which $X_{ij}$ is the squared age of case $j$ at the time that case $i$ is decided. We include these covariates because \citet{black2013citation} find that the number of citations to a Supreme Court case over time depends significantly on the age of the case, characterized by a sharp drop off and leveling out with age. 

\citet{benjamin2012standing} study the propensity for cases to be overruled and cited in other negative ways. They find that cases with majority coalitions that are large and ideologically broad are less likely to be cited negatively. In our data we do not differentiate between negative and positive citations, but since the overwhelming majority of citations are positive, we hypothesize that the effects they found will be reversed in our analysis. We include one covariate in which $X_{ij}$ is the number of justices in the majority coalition for case $j$. We also include another covariate in which $X_{ij}$ is the absolute difference between the maximum and minimum ideal points of the justices in the majority coalition for case $j$. We expect these covariates to have positive effects.

The final control variable we include in the model is one in which $X_{ij} = 1$ if case $j$ was overruled prior to the term in which case $i$ was decided. This variable, quite simply, models the effect of being overruled on the rate of citation to a case after the overruling citation. \citet{fowler2008authority} find that citations to a case drop off quickly after the case has been overruled.

\subsubsection{Dependence terms}\label{dependence_terms}

We include dependence terms to account for each of the dynamics that we hypothesize will characterize the Supreme Court Citation Network---transitivity, reciprocity, and popularity.  The common dependence effects for these three dynamics are the number of triangles, the number of mutual dyads, and the number of in-2-stars. A triangle is a configuration in which there are citations connecting all three cases in a triad. From the perspective of a single tie closing a triad, a positive triangle coefficient indicates that a tie is more likely to form between nodes $i$ and $j$ if nodes $i$ and $j$ have citation ties with one or more cases $k$. A mutual dyad is a configuration in which two nodes within a dyad exchange directed ties. From the perspective of a node closing a mutual dyad with a tie, a positive coefficient indicates that case $i$ is more likely to cite case $j$ if case $j$ has cited case $i$ (mutual dyads are only possible within the same term). The in-2-star configuration is one in which two nodes in a triad send ties to the same third node. From the perspective of a case sending a tie to close an in-2-star, a positive coefficient indicates that case $i$ is more likely to send a tie to case $j$ if one or more other cases also cite case $j$. Unfortunately, adding the triangle or in-2-star statistic causes model degeneracy for the Supreme Court Citation Network---a common issue with these terms \citep{Handcock.2003}. Model degeneracy is a common obstacle when fitting ERGMs, and occurs when the model places most of its probability mass on just a few networks, usually the empty and the full network. In this scenario, the simulated networks are too different from the observed network, making the underlying distribution defined by the model extremely unrealistic.

The statistic included in the model for reciprocity counts the number of mutual dyads (i.e., dyads in which cases $i$ cites case $j$, and case $j$ cites case $i$) in $C_{\leq t}$. In practice, this is the number of mutual dyads in $C_{\leq t}$, since mutual dyads must emerge among opinions that were written in the same term. The log odds interpretation of the reciprocity statistic is that if the opinion in case $i$ cites case $j$, the log odds that case $j$ also cites $i$ change by the estimated coefficient relative to the configuration in which case $j$ does not cite case $i$. We expect the reciprocity effect to be positive.

Accounting for the popularity effect through the in-2-star statistic is prone to cause model degeneracy. In an attempt to stabilize the model against model degeneracy as well as to still capture the popularity effect, \citet{SnijdersTomA.B..2006} introduces the \textit{alternating-in-k-star} statistic, which was shown to be equivalent to the \textit{geometrically weighted indegree distribution} (gwidegree) statistic introduced by \citet{hunter2006inference}. The indegree distribution is the distribution of the number of ties sent to nodes, across all nodes in the network. Define $D_r(C_t, C_{<t})$ as the number of nodes with indegree $r, r \in \{0, \dots , N_t -1 \}$, where $N_t$ is the number of cases in the network at time $t$, then gwidegree represents a network's indegree distribution in a single statistic by geometrically weighting the degree distribution
\begin{equation}
h_{gwidegree}(C_{t}, C_{<t}, \lambda)= \lambda \sum_{r=1}^{N_t-1}\biggl(1-\bigl(1-\frac{1}{\lambda}\bigr)^r\biggr)D_r(C_t, C_{<t}). 
\label{gwidegree}
\end{equation}
The decay parameter $\lambda \in [0, \infty ]$ controls the decline of the weight put on each node in the network based on their indegree ($D_r(C_t, C_{<t})$) as the indegree value ($r$ ) increases, which means that the higher $\lambda$ the more the statistic weighs cases with a high indegree. We fixed $\lambda=1$. We chose this value for two reason. First, it results in a fairly straightforward interpretation of the coefficient for gwidegree, as Equation \ref{gwidegree} reduces to a function that counts the number of nodes that receive at least one tie. Our second reason for selecting $\lambda=1$ is that it results in an accurate fit in the indegree distribution of the observed network (as demonstrated in the appendix). At $\lambda=1$, the weight attributed to each case is equivalent regardless of the indegree value, as long as the indegree is greater than zero. This means that the statistic does not grow with the addition of high indegree cases to the network. It only grows through the addition of cases with at least one in-citation. With  $\lambda=1$, the coefficient associated with the gwidegree statistic governs the degree to which the edges are sent to a small number of popular nodes, or distributed fairly evenly across the nodes. A positive coefficient encourages edges to be distributed evenly across a large number of nodes---assuring that as many nodes as possible receive at least one tie. A negative gwidegree coefficient discourages network configurations in which many nodes receive ties, placing higher probability instead on configurations in which many of the ties are sent to a relatively small number of popular nodes. As stated by \citet{Levy2016}, this means for the interpretation that a negative parameter value indicates the centralization of edges (i.e., popularity) while a positive parameter indicates the dispersion of edges (i.e., new ties going to less popular nodes).  Since we expect highly cited cases to be more likely to be cited again, we expect the gwidegree parameter to be negative.

We include two statistics to account for different types of transitivity. The first transitivity statistic calculates the number of transitive ties, $i \to j$, where there is a directed path of length two from $i$ to $j$ through at least one case $k$, and $j$ and $k$ were written during a different term than $i$. We will refer to this statistic as the \textit{different term transitivity} statistic. This statistic captures the central form of transitivity that we hypothesize above---when a justice writes the opinion for case $i$, cites a past opinion ($k$) as precedent, and then cites one or more of the opinions ($j$) that were cited in opinion $k$. The two-path from opinion $i$ to $j$ is created when opinion $i$ cites an opinion ($k$) that cites an earlier opinion ($j$). The closed transitive triangle is created when opinion $i$ cites directly to $j$, one of the precedents used in opinion $k$. The edge $i \to j$ counts as a ``transitive tie'' since it closes at least one transitive triangle. The \textit{different term transitivity}  statistic is the count of the number of transitive ties in the network that connect opinions written in different terms. As illustrated in Section 2, we theorize Supreme Court citations to exhibit a high degree of transitivity. As a result, we expect positive parameter estimates for the \textit{different term transitivity} statistic.

Whereas the first transitivity statistic is focused on modeling triangles that form through the citation of past cases, the second transitivity statistic we include captures clustering that includes ties formed between cases decided in the same term. We both expect and observe a higher level of transitivity among same-term cases since justices have the opportunity to confer about cases and coordinate citations among related opinions. For modeling within-term transitivity, we use the \textit{geometrically weighted edgewise shared partners} (gwesp) statistic introduced by \citet{hunter2006inference}. We use the form of the gwesp statistic that captures configurations in which cases $i$ and $j$ are connected by a citation ($i$ cites $j$, $j$ cites $i$, or both), and also cite $r$ common other cases (i.e., shared partners). \citet{Butts.2008} call this conceptualization of $r$ the number of \textit{outgoing shared partners}. This form of gwesp captures the degree to which two related cases in the same term are likely to be based on similar bodies of precedent. If the gwesp coefficient is positive, we interpret the effect as saying that the more shared partners $r$ to which cases $i$ and $j$ send ties, the greater the likelihood that there is a citation connecting cases $i$ and $j$.  As with the gwdegree term, there is a decay parameter. The gwesp statistic exhibits decreasing marginal returns with respect to the number of shared partners (e.g., the fourth shared partner between $i$ and $j$ does not contribute as much to gwesp as does the first shared partner between $i$ and $j$). The decay parameter governs the degree of marginal return decrease, with a value of 0 indicating that additional shared partners do not contribute anything, and a value of $\infty$ indicating a linear (non-decreasing) function.  We fix the decay parameter at $\lambda=0.25$, as this accurately captures the edgewise shared partner distribution, and avoids degeneracy (which would be a problem at higher values of $\lambda$). In general, researchers can adjust the gwesp parameter down to avoid degeneracy, and increase it to more accurately capture the shared partner distribution in the network. As with cross-term transitivity, we expect the gwesp statistic to also carry a positive coefficient value.

We include one more structural network effect to adjust for inherent differences across cases in terms of the overall scope of the legal issues addressed in a case.  \citet{fowler2008authority} finds that cases vary in the degree to which they serve as ``hubs''---citing to many other cases. We expect that new cases will be more likely to cite cases that themselves cited many cases, because cases that themselves cite a large number of cases address a broader array of legal issues. We include the outdegree (i.e., the number of citations sent) of every node that entered the network prior time $t$ as a receiver effect. We expect the coefficient on this term to be positive, as a positive coefficient would mean that cases that send more citations are likely to also receive more citations. This is not technically a dependence term, as it is formulated as a covariate, but we discuss it in the current section because it is purely a network structure effect.

\begin{figure}[bt]
	\centering
	\includegraphics[width=10cm ]{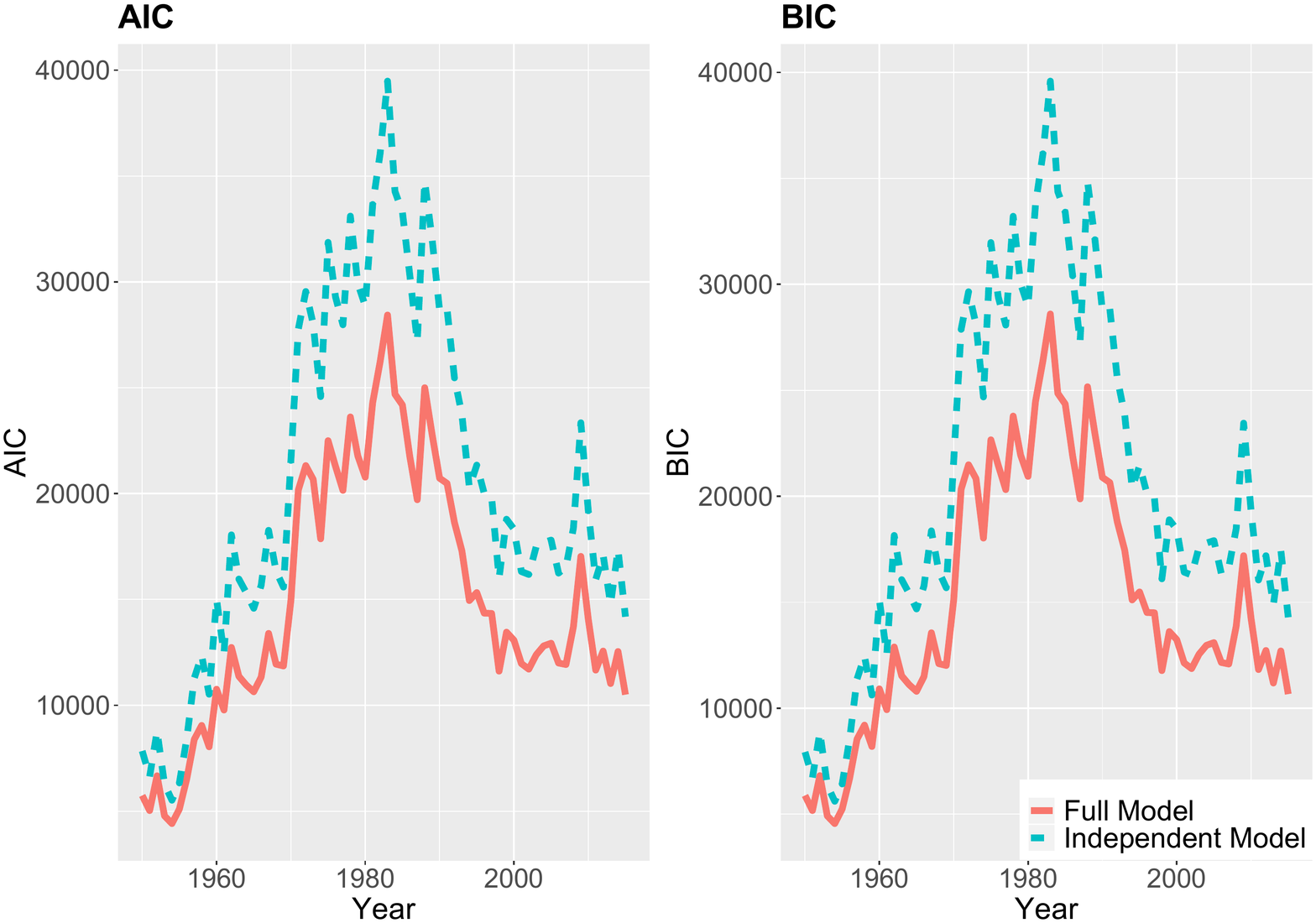}
	\caption{AIC and BIC for the full and the independent model for the time frame 1950-2015 }
	\label{AIC_BIC}
\end{figure}

\subsection{Model Fit}

Our case for studying legal citations at the directed dyadic level hinges upon the contribution to modeling offered by incorporating network dependence. To quantify this contribution, we fit one model that incorporates covariate effects only, which we term the {\em Independent Model}---excluding the reciprocity, transitivity, and popularity terms. The full model is the one we present below. We use both AIC and BIC to compare the fit of the two models, with lower values indicating better fit.

Figure \ref{AIC_BIC} depicts the AIC and BIC comparison for the two different models for each term from 1950 to 2015. We see that the AIC and BIC is considerably smaller for the full model throughout the period of consideration. On average the independent model yields a 5,525 point higher AIC and an average 5,582 point higher BIC. The maximum difference is reached in 1983 with a difference of 10,990 points for the AIC and 10,930 points for the BIC. This fit comparison provides robust evidence that the dependence terms contribute significantly to the explanatory power of the model. In the appendix we present conventional goodness-of-fit plots that are used to assess the structural fit for ERGM \citep{hunter2008goodness}, as well as degeneracy diagnostics \citep{mukherjee2020degeneracy}. We find that the full model fits the degree and edgewise shared partner distributions well, and is not degenerate.

\subsection{cERGM Coefficient Estimates}
The estimates from the cERGM are presented in Figures \ref{SCC_results_1} and \ref{SCC_results_2}. In these figures we depict coefficient estimates and 95\% confidence intervals for all of the effects for all the variables we discussed in section \ref{covariate_terms}. The coefficients in an ERGM family model can be interpreted in the same way as logistic regression coefficients---they give the change in the log odds of a tie from $i$ to $j$ given a one-unit increase in the respective variable. In our figures, a circle indicates that the estimate was significant at an $\alpha=0.05$ level, while an square translates to a p-value between $0.05$ and $0.1$. A triangle signals that the estimate for a given year was not significant at an $\alpha=0.1$ level. In each panel, the background shading indicates different chief justice courts, with the two darker shades representing, respectively, the Warren (1954-1969) and Rehnquist (1987-2005) courts. Finally, note that the panels showing results of the reciprocity statistic and the \textit{Overruled Cases} variable do not provide estimates for every term. In these missing terms there were no two cases that cited each other or no overruled cases being cited, making the estimate of these terms equivalent to $-\infty$. 

\subsubsection{Dependence Terms}
\begin{figure}
	\centering
	\includegraphics[width=14cm ]{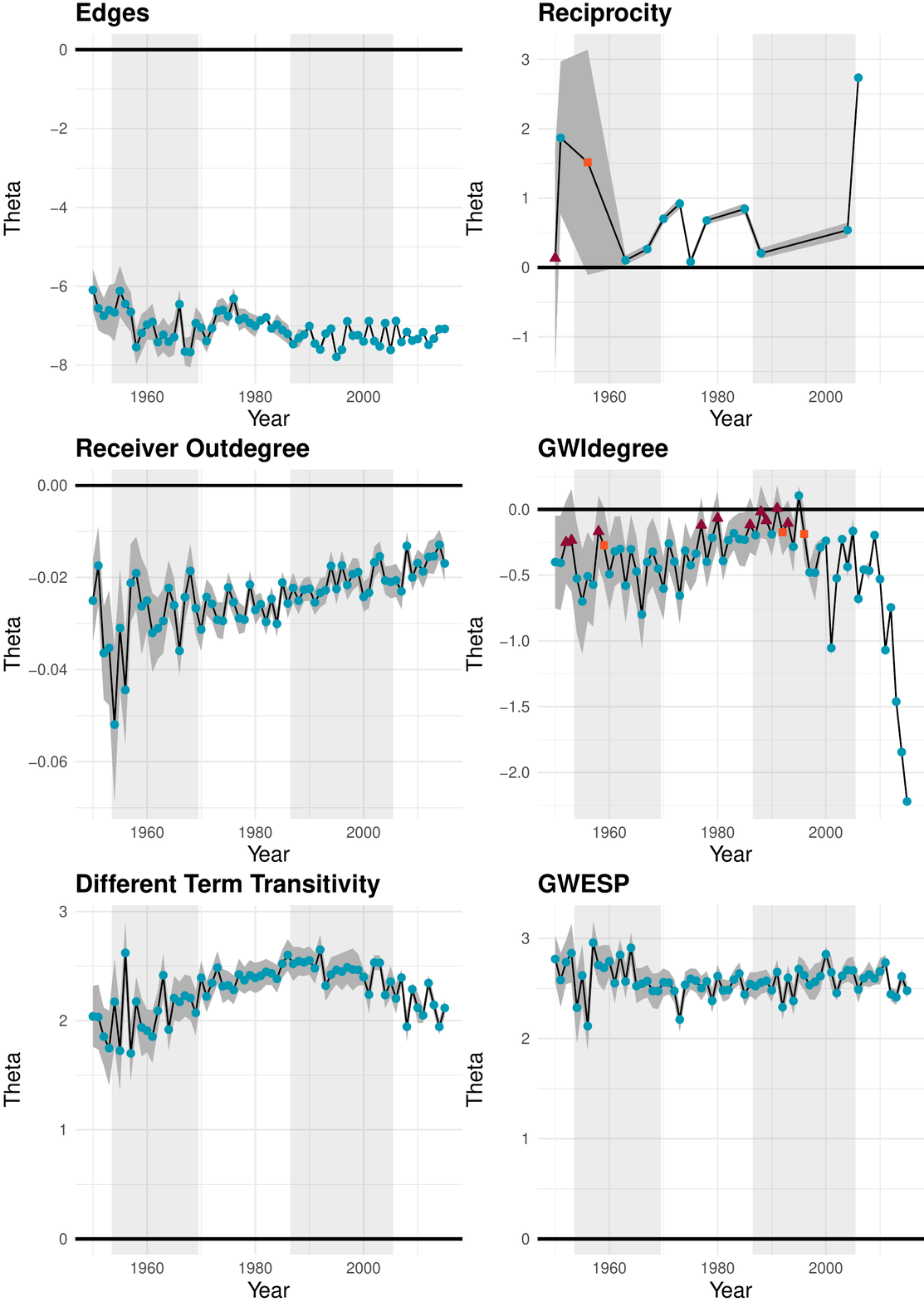}
	\caption{ERGM results for the dependence terms. Circles indicate a p-value smaller than 0.05, squares a p-value between 0.05 and 0.1 and triangles a p-value greater than 0.1. Different chief justice terms are indicated by shading in the background; the two grey areas indicate the Warren and Rehnquist courts.}
	\label{SCC_results_1}
\end{figure}

We first discuss the dependence effect estimates in Figure \ref{SCC_results_1}. In most of the years that it is estimable, the reciprocity effect is statistically significantly positive, and substantial in magnitude. In a typical year, the presence of a citation from case $i$ to case $j$ increases the log odds of a citation from case $j$ to case $i$ by approximately 0.50. Both transitivity effects---different term transitivity and gwesp---are estimated to be positive and statistically significant in every year, and are even greater in magnitude than the reciprocity effect. The log odds of a citation from case $i$ to case $j$ increases by more than 2 if there is at least one third case, $k$ that is cited by $i$, and cites to $j$. Results regarding the popularity effect (i.e., gwidegree) are generally supportive of the hypothesis of a popularity effect in Supreme Court citations, as the gwidegree estimate is negative and statistically significant at conventional levels in nearly every term.

Contrary to expectations, we find that the hub effect is negative. That is, each additional citation sent by a case reduces the likelihood that the case is cited. For every citation sent, the log odds of any citation to a case decreases by an average of approximately 0.025---a result that is statistically significant in each term. We initially expected receiver outdegree to have a positive effect because opinions with broad scopes can be useful citation hubs. Instead, our finding indicate that high outdegree more importantly reflects case characteristics that negatively impact incoming citations. One likely characteristic is case complexity. Prior studies show that justices tend to take a more balanced approach when faced with complex cases \citep{collins2008amici, lindquist2007splitting}, which can yield opinions that cite widely \citep{wilkinson2005rehnquist}. At the same time, it has been shown that complex cases are more likely to be overruled \citep{spriggs2001explaining}. To the extent, then, that outdegree count captures case complexity, our finding suggests that justices cite more direct alternatives in anticipation of complex opinions being overruled.

The negative effect of receiver outdegree, which seems to result from strategic anticipation by the justices, ostensibly decreases with time. Specifically, term to term variation aside, there is an upward trend toward zero in the parameter estimates. We believe this finding is particularly illustrative of why our approach of fitting separate models for each term is appropriate. As time passes, the number of nodes in the network (i.e., cases) grows. With the increasing number of citable cases and a relatively stable baseline rate of citation (i.e., the edges effect), there is an increase in outdegree mean and variation. This increasing difference between outdegree count by case makes it unlikely for the per unit change effect in outdegree to remain at a constant level. We have then an instance of network growth leading to changes in the model parameters even if the underlying network process remains unchanged. Approaches that cannot encompass these complexities are likely to produce poor fitting models.

\subsubsection{Covariate Terms}
\begin{figure}
	\centering
	\includegraphics[width=14cm ]{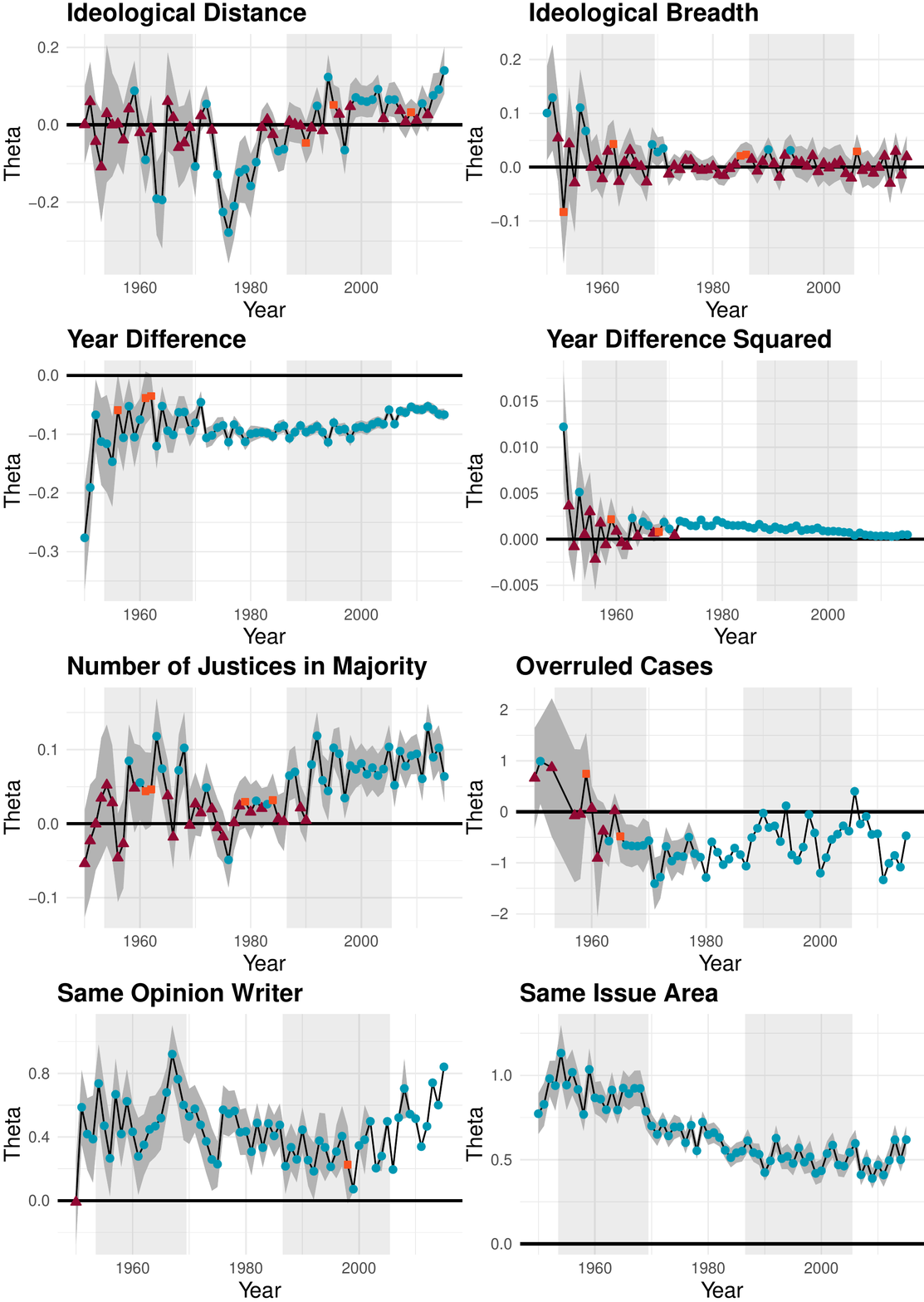}
	\caption{ERGM results for the covariate terms. Circles indicate a p-value smaller than 0.05, squares a p-value between 0.05 and 0.1 and triangles a p-value greater than 0.1. Different chief justice terms are indicated by shading in the background; the two grey areas indicate the Warren and Rehnquist courts.}
	\label{SCC_results_2}
\end{figure}

Effects of the exogenous covariates included in the cERGM are presented in Figure \ref{SCC_results_2}. Most of the exogenous covariate effects align with expectations and have relatively stable parameter estimates over time. We find that cases are more likely to cite those that (1) have been decided most recently, (2) were authored by the same justice, (3) are in the same issue area, and (4) have not been overruled. Surprisingly, we do not find consistent effects for majority coalition size or for the two covariates that involve justice ideology. The effect of majority coalition size was generally not statistically significant until the later portion of the period examined; ideological distance between two cases had signs and significance levels that are highly inconsistent; and the effect of the ideological breadth of the potentially cited case was effectively zero for the entire period. 

Our results leave open the question of whether and how citations are shaped by ideological factors. A purely measurement-based explanation of our findings is that we need a more precise measure of the ideological position of an opinion, as far as citation behavior goes, than the median ideal point of the justice in the majority coalition. Clustering that results from ideological homophily that is not effectively accounted for by our measure would be picked up, in part, by our measures of transitivity, which are statistically and substantively significant throughout the period studied. 

Beyond potential measurement issues, we do observe connections between our findings and the literature on decision-making on the Supreme Court. These connections are evident from a closer look at temporal dynamics in the effects. In this discussion, we focus on the shift in the effect of ideological distance beginning with the Rehnquist court and extending into the Roberts court. 

The existing body of evidence indicates that chief justices impact the courts they preside over beyond casting one of the nine votes \citep{cross2005decisional,danelski2016chief}. It is therefore within expectations that there are discernible differences in citation patterns across different courts. With regard to the effect of ideological distance, we see that the Vinson (1950-1953) and Warren (1954-1969) courts were highly volatile, the Burger court (1970-1986) had a generally higher likelihood of citing opinions that were ideologically closer to itself, and the more recent Rehnquist (1987-2005) and Roberts (2006-2015) courts showed some tendency toward citing cases that are ideologically distant. The latter finding might be initially puzzling, but can be explained in light of observations from legal scholars that the Rehnquist court was characterized by a `split-the-difference' jurisprudence toward its later half \citep{wilkinson2005rehnquist,basiak2006roberts}. This means that the court strove to take moderate positions that compromised on both sides of the debated issues, which in large differed from the jurisprudential approach of prior courts. Less has been written about the Roberts court on this topic, but Roberts shares many similarities with Rehnquist especially in comparison to previous chief justices, and has himself noted the desirability of widely accepted rulings even at the expense of their scope \citep{pomerance2018center,sunstein2008trimming}.

One way to achieve the kind of balance required for support is to cite broadly from both sides of the argument \citep{wilkinson2005rehnquist}, which increases the likelihood of citing ideologically distant cases. Consider for example Hamdi v. Rumsfeld (542 U.S. 1, 2004) which ruled on the question of whether a U.S. citizens can be detained as an ``enemy combatant''. The court, in a four-justice plurality opinion supported by two additional concurring justices, decided at the same time that ``a state of war is not a blank check for the President'' and the due process assessment must ``pay keen attention to the particular burdens faced by the Executive in the context of military action''. In total, the plurality opinion in Hamdi v. Rumsfeld cited 47 opinions, with cites as distant as 2.42 and 2.13 on either side of its median position.

In the above discussion, we identified an explanation for the observed temporal dynamics in the effect of ideological distance. Specifically, ideologically distant cases become more important as courts adopt the relatively new `split-the-difference' jurisprudential approach which requires broad citations. We recognize that there are alternative explanations---perhaps one that is related to the coterminous change in the effect of majority coalition size---but a deeper look into the relevant literature is beyond the scope of this paper. Our discussion serves to illustrate that term-by-term cERGMs can uncover important temporal dynamics that courts researchers can use to study a much greater range of citation-related phenomenon beyond what we present here. To the extent that the US Supreme Court is a political entity that interacts with the broader society, the body of court opinions and citations between them is corpus that continues to grow based on both its history and its contemporary context. The methods we present here afford researchers the tools to better understand these complexities.

\section{Conclusion}

We present a methodology for studying citations between US Supreme Court opinions at the dyadic level, as a network. This methodology---the citation ERGM---enables researchers to include both exogenous covariates such as the ideological predisposition and age of a case, and dependence terms, such as transitivity and reciprocity, as explanations for citation formation. We apply this methodology to a network that includes all Supreme Court cases decided between 1950 and 2015. We find, somewhat counterintuitively, that Supreme Court citations are highly reciprocal. We also find that citations are driven by dependencies such as triad closure and popularity. The dependence effects that we identify are as substantively and statistically significant as the effects of the exogenous covariates we include in the model. The summary result from this analysis is that theoretical models of Supreme Court citation formation should consider both the effects of case characteristics and the structure of past citations. 

Though we see the advancements in modeling the Supreme Court citations as our central contribution, we make two contributions to broader literatures on judicial citations and citation analysis more generally. First, our arguments regarding the dependencies that shape Supreme Court citations are likely to apply to citation networks formed among other court opinions, and we have provided a road map and tools for modeling such dependencies. Second, we have provided an extended version of ERGM that is appropriate for all forms of citation network analysis, and is available in a convenient statistical software package.

\singlespacing
\bibliography{bib} 
\bibliographystyle{apsr}

\onehalfspacing
\appendix
\section{cERGM Estimation}

The normalizing constant in Equation 1 of the main text is intractable. For example, in the simple case of adding three cases to a network in which six cases already exists---like that depicted in Figure 4 of the main text---there are 16,777,216 unique configurations of $C_t$ that could be observed. The typical Supreme Court term involves adding hundreds of cases to a network that already includes thousands of previous cases. This means that straightforward methods of maximum likelihood estimation (MLE) are infeasible with the cERGM. 

The common alternative relies on Monte Carlo methods to approximate the normalizing constant by simulating a large set of networks \citep{hunter2006inference,hummel2012improving}. The resulting estimator, the Monte Carlo MLE (MCMLE), is approximately consistent, meaning that it converges to the MLE as the sample size, i.e., the number of simulated networks, increases. However, one drawback is that with the number of nodes in the Supreme Court citation network being in the order of $10,000$, obtaining the MCMLE is computationally expensive \citep{schmid2017exponential} and the success of the algorithm heavily relies on the starting parameter vector $\theta_0$, which is ideally chosen in the proximity of the unknown MLE \citep{hummel2012improving}. The prevailing choice for $\theta_0$ is the maximum pseudo-likelihood estimation (MPLE) \citep{strauss1990pseudolikelihood}, a fast estimation method that is defined as maximizing the log product of the conditional probability of each citation (and non-citation), conditional on the other elements of the observed citation network. The joint probability of all citations is replaced by the product over conditional probabilities, which, as we demonstrated in the main text, assume a logit form. The MPLE is simple to obtain, but does not guarantee a starting value close to the MLE \citep{SchmidHunter2020}. 

The MCMLE of networks up until the 90s was obtainable in a reasonable time frame starting at the MPLE and sampling $10,000$ networks to approximate the normalizing constant. However, the estimation of most networks in the 90s with the MPLE as starting values was not feasible in an reasonable time frame anymore. Instead, we improved the choice of starting value $\theta_0$ by fixing it at the MCMLE of the previous term $t-1$ and successfully obtained the MCMLE of the network at term $t$. But even this approach started to fail for networks around the turn of the millennium. Neither the MPLE nor the MCMLE of previous terms as starting values led to successful estimation, and neither did the Stepping algorithm \citep{hummel2012improving}. 
The MCMLE for these large citation networks was obtained by setting the starting value according to an novel approach introduced by \citet{SchmidHunter2020}. This method is based on the fact that the MLE of exponential family distributions is solely a function of the vector of sufficient statistics $h(C_t, C_{<t})$, meaning that the MLE of two networks $A$ and $B$ is equal if $h(A)=h(B)$. However, the MPLE of networks with the same sufficient statistics is not necessarily the same. Instead of starting the MCMLE algorithm at the network's MPLE, \citet{SchmidHunter2020} propose searching for a new network $C_t^*$ on the same nodes as the observed network that satisfies $h(C_t, C_{<t}) = h(C_t^*, C_{<t})$, and has a weak dependence structure among unfixed ties. Such a network can be found using simulated annealing algorithms \citep{Kirkpatrick83}. For networks with a weak dependence structure among unfixed ties, the MPLE is similar to the MLE, in addition, the same sufficient statistic between $C_t^*$ and the observed network $C_t$ guarantees the same MLE between these two networks. This makes the MPLE of $C_t^*$ an effective starting value for the MCMLE algorithm. Since for some networks the MCMLE was only obtainable using simulated annealing method to find a starting value, the final results in the paper have all been estimated using simulated annealing.
The simulated annealing algorithm for finding an improved starting value for cERGMs was implemented in the \textbf{cERGM}-package for \textbf{R} \citep{RCore} and can be found at \url{http://github.com/schmid86/cERGM}.

\section{Goodness-of-Fit}
\begin{figure}[bt]
	\begin{center}
		\includegraphics[width=0.9\textwidth]{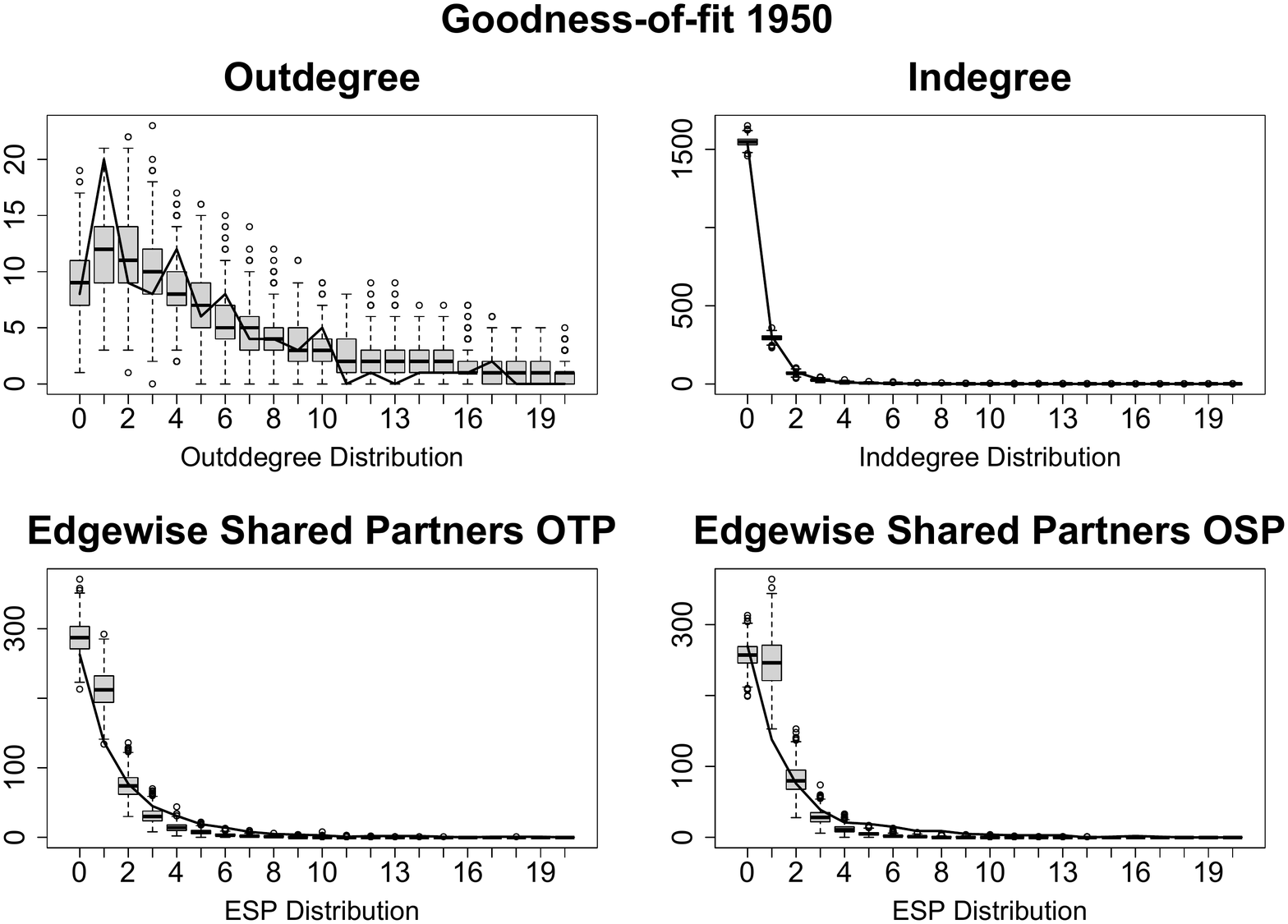}\\
		\vspace{1cm}
		\includegraphics[width=0.9\textwidth]{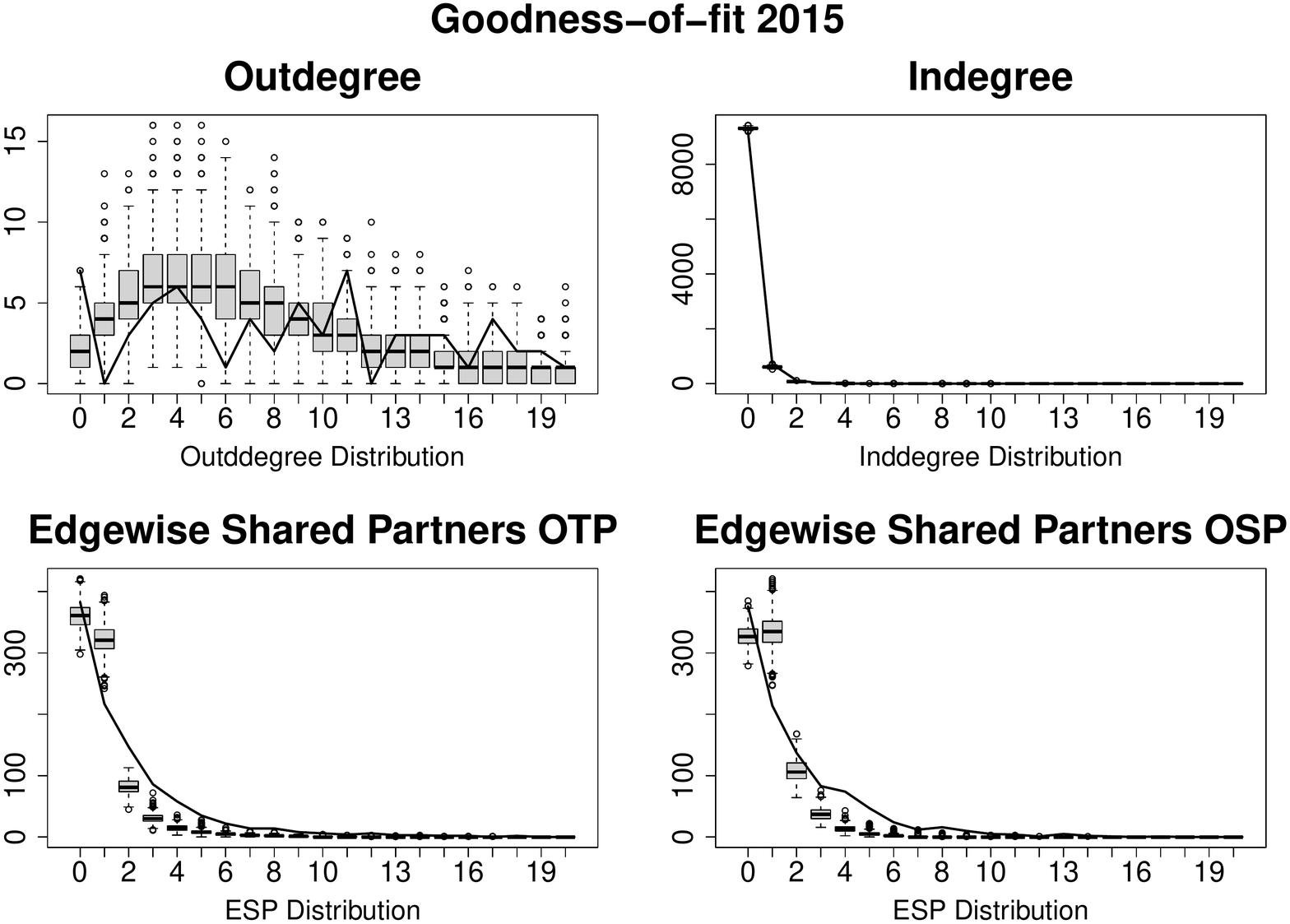}
		\vspace{0.1cm}
		\caption{Goodness-of-fit diagnostic for the 1950 network (top) and the 2015 network (bottom).}
		\label{GOF}
		\vspace{-.25cm}
	\end{center}
\end{figure}

We evaluate the goodness-of-fit of the model following \citet{hunter2008goodness} by examining the distribution of four hyper statistics, e.g., the out- and indegree distribution and the distribution for two different edgewise shared partners statistics. OTP stands for \textit{outgoing two-paths} and refers to the number of cases $r$ that are cited by case $i$ and that cite case $j$, while $j$ is also directly cited by $i$. The second ESP statistic is the OSP specification that has been introduced in section 4.1.2 in the paper. Figure \ref{GOF} visualizes the goodness-of-fit results for the citation network for the 1950 (top) and 2015 (bottom) term. The solid black line indicates the statistic's distribution in the Supreme Court citation network of that given term and the boxplots depict the statistic's distribution of $1000$ networks that have been simulated from the ERGM defined by the MCMLE. This means that in the ideal case the solid black line passes through ever single boxplot.

We see that our models do a good job capturing the out and indegree distribution of the citation network, since the black line falls almost exclusively within the ranges spanned by the boxplots. For the ESP distributions we can observe that the number of ties with $r=0$ shared partners is captured well for both the OTP as well as for the OSP statistic. However, the model overestimates the number of $r=1$ shared partners and then, especially in the 2015 term network, underestimates the number of ties with more than $r=1$ shared partners.

\section{Checking for Model Degeneracy}

\begin{figure}[bt]
	\begin{center}
		\includegraphics[width=1\textwidth]{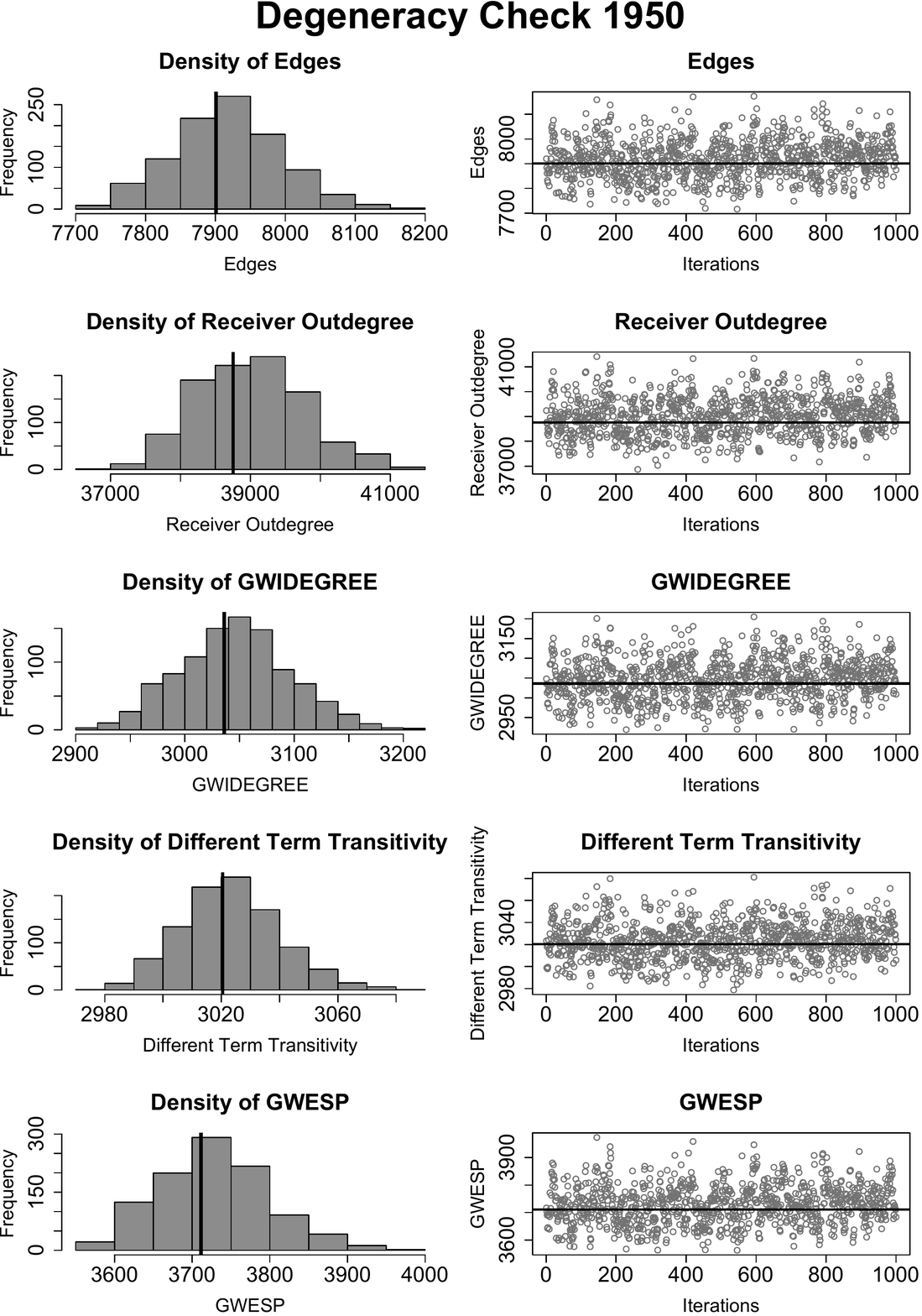}
		\caption{Density and trace plots for the dependency terms of the 1950 term citation network.}
		\label{mcmcdiagnostics_1950}
		\vspace{-.25cm}
	\end{center}
\end{figure}

\begin{figure}[bt]
	\begin{center}
		\includegraphics[width=1\textwidth]{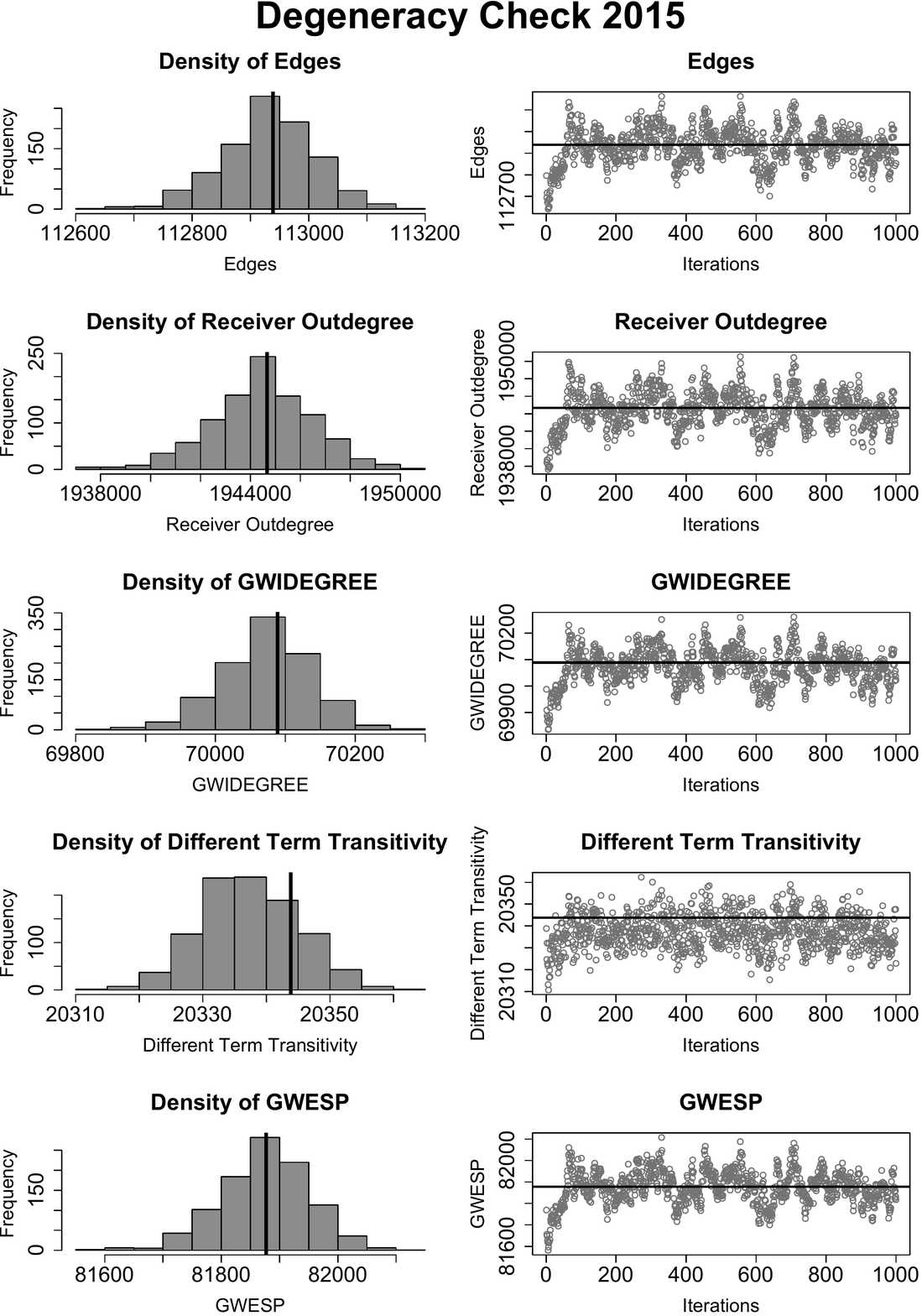}
		\caption{Density and trace plots for the dependency terms of the 2015 term citation network.}
		\label{mcmcdiagnostics_2015}
		\vspace{-.25cm}
	\end{center}
\end{figure}

A common challenge when fitting ERGMs is model degeneracy. Model degeneracy occurs when the probability distribution defined by the parameter vector does not predominantly yield networks with similar statistics as the observed network. Generally, model degeneracy results in simulated networks with no ties or all possible ties. In a non-degenerate model the statistics of the networks that were simulated from the probability distribution defined by the MCMLE fall in the proximity of the observed network's statistics. Figures \ref{mcmcdiagnostics_1950} and \ref{mcmcdiagnostics_2015} depict trace and density plots for the dependence terms in the 1950 and 2015 term citation network. The histograms on the left visualize a statistic's density from $1000$ simulated networks, while the right side shows the statistic's trace plot of the same $1000$ networks. The solid black line indicates the statistic's value in the actual citation network. Both figures indicate that this model is non-degenerate and that the simulated network's statistics fall almost evenly around the observed statistic. The density and trace plots for the ERGM of the terms not depicted provide similar results.

\end{document}